\documentclass[runningheads]{llncs}

\usepackage[utf8]{inputenc}
\usepackage{CJKutf8}
\usepackage{eccv}
\usepackage{fancyhdr}

\usepackage{makecell}
\usepackage{colortbl}
\usepackage{booktabs}
\usepackage{multirow}
\usepackage{tabularx}
\usepackage[table]{xcolor}

\usepackage{eccvabbrv}

\usepackage{graphicx}
\usepackage{booktabs}
\usepackage[accsupp]{axessibility}  
\usepackage{eccvabbrv}
\usepackage{graphicx}
\usepackage{booktabs}
\usepackage[breaklinks,colorlinks,citecolor=eccvblue]{hyperref}
\usepackage{orcidlink}
\usepackage{multirow} 
\usepackage{tcolorbox}
\tcbuselibrary{skins, breakable}
\usepackage{tcolorbox}
\tcbuselibrary{skins, breakable}
\usepackage{enumitem}

\usepackage{booktabs}
\usepackage{multirow}
\usepackage{amssymb}
\usepackage{pifont}
\usepackage{xcolor}
\usepackage{colortbl}
\usepackage{graphicx}
\usepackage{enumitem}
\usepackage[table]{xcolor}
\usepackage{booktabs}
\usepackage{multirow}
\usepackage{graphicx} 
\usepackage{amsmath}  

\newcommand{\cmark}{\ding{51}}
\newcommand{\xmark}{\ding{55}}
\definecolor{highlightgray}{gray}{0.95}
\definecolor{omninavy}{HTML}{102A72}
\definecolor{omniteal}{HTML}{1A7F7A}
\definecolor{omnigray}{HTML}{707784}

\newcommand{\omniapplypagestyle}[2]{%
  \fancyhf{}%
  \fancyhead[LE,LO]{\textcolor{omninavy}{\footnotesize\sffamily\textbf{#1}}}%
  \fancyhead[RE,RO]{\textcolor{omnigray}{\footnotesize\sffamily #2}}%
  \fancyfoot[LE,LO]{\textcolor{omnigray}{\footnotesize\sffamily XPeng Motors}}%
  \fancyfoot[RE,RO]{\textcolor{omniteal}{\footnotesize\sffamily\textbf{\thepage}}}%
  \renewcommand{\headrulewidth}{0.7pt}%
  \renewcommand{\footrulewidth}{0.4pt}%
  \renewcommand{\headrule}{\hbox to\headwidth{\color{omninavy}\leaders\hrule height \headrulewidth\hfill}}%
  \renewcommand{\footrule}{\hbox to\headwidth{\color{omniteal}\leaders\hrule height \footrulewidth\hfill}}%
}

\fancypagestyle{omniguistyle}{%
  \omniapplypagestyle{OmniGUI}{}%
}

\fancypagestyle{plain}{%
  \omniapplypagestyle{OmniGUI}{}%
}

\makeatletter
\let\ps@headings\ps@omniguistyle
\let\ps@myheadings\ps@omniguistyle
\makeatother

\begin{document}

\title{OmniGUI: Benchmarking GUI Agents in Omni-Modal Smartphone Environments}

\titlerunning{}

\author{Felix Henry\textsuperscript{*}\inst{1} \and Lihan Lin\textsuperscript{*}\inst{1} \and Jiangyou Zhu\inst{1} \and Yangfan\inst{1} \and Bingqian Zhang\inst{1} \and Min Chen\inst{1} \and Shiyu Huang\textsuperscript{\ensuremath{\dagger}}\inst{1}}

\authorrunning{}

\institute{XPeng Motors\\
\textsuperscript{*}Equal contribution. \quad \textsuperscript{\ensuremath{\dagger}}Corresponding author.}

\maketitle
\pagestyle{omniguistyle}
\thispagestyle{omniguistyle}

\begin{abstract}
Current benchmarks for graphical user interface (GUI) agents predominantly rely on static screenshots. However, real-world smartphone interaction routinely requires agents to process transient audio cues and temporal video dynamics that are tightly coupled with the moment of action. To bridge this gap, we introduce OmniGUI, the first step-level benchmark designed to evaluate GUI agents in omni-modal smartphone environments. OmniGUI provides continuous, interleaved multimodal inputs---comprising static images, synchronous audio, and video clips---at every action step. The dataset encompasses 709 expert-demonstrated episodes (2,579 action steps) across 29 applications, systematically annotated with objective multimodal dependency levels. Because dedicated omni-modal GUI agent frameworks are currently in their nascent stage, we select foundational omni-modal models capable of natively processing interleaved inputs to serve as agent proxies for our initial baselines. Our empirical evaluation reveals that while current models exhibit competency on visually static tasks, their action prediction performance degrades significantly in environments requiring synchronous temporal and auditory signals. Furthermore, ablation studies isolate specific operational bottlenecks, notably cross-modal interference when processing task-irrelevant environmental noise. The complete dataset, evaluation pipeline, and baseline prompts are provided in the supplementary material. Project page: \href{https://omni-gui.github.io}{https://omni-gui.github.io}.

\end{abstract}

\begin{figure*}[t]
\centering
\includegraphics[width=\textwidth]{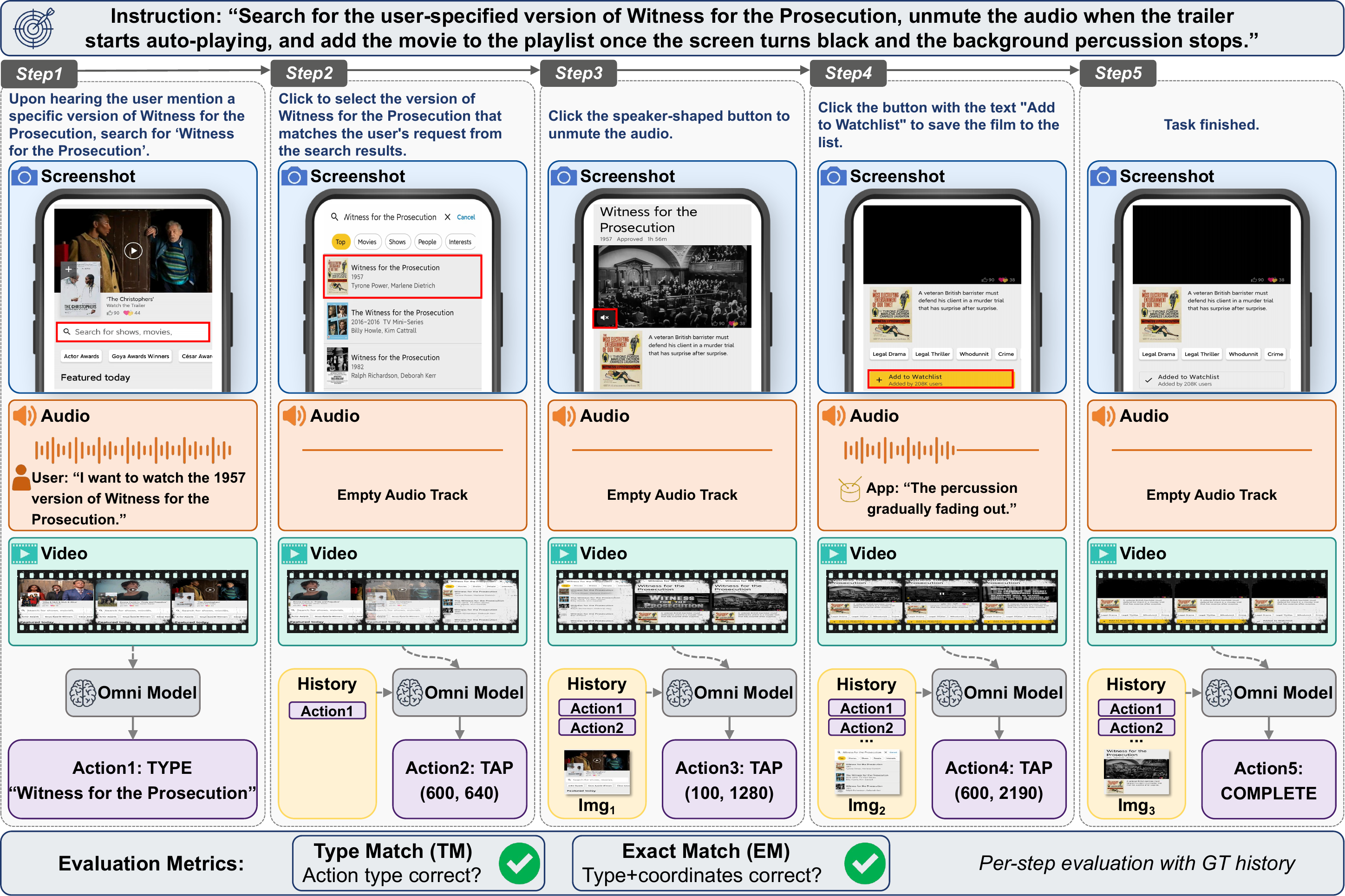}
\caption{
\textbf{Overview of the OmniGUI benchmark framework.} 
Given a multimodal instruction (e.g., a spoken request), a GUI agent interacts with a smartphone interface across multiple steps. At each action step, the agent receives interleaved inputs comprising a static screenshot, real-time audio (e.g., user speech, application sounds), and a temporal video clip, along with the action history. Based on these synchronous multimodal signals in the environment, the agent predicts the subsequent action (e.g., \texttt{TYPE}, \texttt{TAP}). Performance is quantitatively evaluated using Type Match (TM) and Exact Match (EM) metrics against ground-truth human demonstrations.
}
\label{fig:teaser}
\end{figure*}

\section{Introduction}
\label{sec:intro}

\begin{table*}[t]
\centering
\caption{
\textbf{Comparison of OmniGUI with representative GUI agent benchmarks.}
\textit{Audio} and \textit{Video} indicate whether the benchmark provides
auditory and video inputs beyond static screenshots.
\textit{Per-Step} denotes whether multimodal inputs are provided at
\emph{every} action step, rather than as pre-task or reference content.
\textit{Action Output} specifies the format of predicted actions.
\textit{Manual} indicates whether all tasks are manually designed.
}
\label{tab:benchmark_comparison}
\vspace{-2mm}
\setlength{\tabcolsep}{4pt}
\renewcommand{\arraystretch}{1.15}
\resizebox{\textwidth}{!}{%
\begin{tabular}{l c r r c c c c c c}
\toprule
\multirow{2}{*}{\textbf{Benchmark}} 
& \multirow{2}{*}{\textbf{Platform}} 
& \multirow{2}{*}{\textbf{\#Tasks}} 
& \multirow{2}{*}{\textbf{\#Steps}} 
& \multicolumn{3}{c}{\textbf{Input Modalities}} 
& \multirow{2}{*}{\textbf{Per-Step}} 
& \multirow{2}{*}{\textbf{Action Output}} 
& \multirow{2}{*}{\textbf{Manual}} \\
\cmidrule(lr){5-7}
& & & & \textbf{Image} & \textbf{Video} & \textbf{Audio} & & & \\
\midrule
\multicolumn{10}{l}{\textit{Vision-Only Benchmarks}} \\
\midrule
AITW~\cite{rawles2023androidinthewild}                     
    & Android & 30,378  & 715K+   
    & \cmark & \xmark & \xmark & \xmark & Coordinate & \xmark \\
GUI-Odyssey~\cite{lu2025guiodyssey}            
    & Android & 7,735   & 74K+    
    & \cmark & \xmark & \xmark & \xmark & Coordinate & \xmark \\
AndroidWorld~\cite{rawles2024androidworld}     
    & Android & 116     & --      
    & \cmark & \xmark & \xmark & \xmark & Coordinate & \cmark \\
Mind2Web~\cite{deng2023mind2web}               
    & Web     & 2,350   & 12K+    
    & \cmark & \xmark & \xmark & \xmark & DOM Element & \xmark \\
OSWorld~\cite{xie2024osworld}                  
    & Desktop & 369     & --      
    & \cmark & \xmark & \xmark & \xmark & Coordinate & \cmark \\
ScreenSpot~\cite{cheng2024seeclick}            
    & Multi   & 1,272   & --      
    & \cmark & \xmark & \xmark & \xmark & Coordinate & \cmark \\
\midrule
\multicolumn{10}{l}{\textit{Benchmarks with Partial Multimodal Support}} \\
\midrule
MM-Mind2Web~\cite{zheng2024gpt}      
    & Web     & 2,000+  & --      
    & \cmark & \xmark & \cmark & \xmark & DOM Element & \xmark \\
GUI-World~\cite{chen2024gui}              
    & Multi   & 12,379  & --      
    & \cmark & \cmark & \xmark & \xmark & QA / Caption & \xmark \\
VideoGUI~\cite{lin2024videogui}                
    & Multi   & 178     & 4K+     
    & \cmark & \cmark & \xmark & \xmark & Coordinate & \cmark \\
VideoWebArena~\cite{jang2024videowebarena}     
    & Web     & 2,021   & 24K--38K 
    & \cmark & \cmark & \cmark & \xmark & DOM Element & \xmark \\
\midrule
\multicolumn{10}{l}{\textit{Per-Step Multimodal Benchmark}} \\
\midrule
\rowcolor{blue!8}
\textbf{OmniGUI (Ours)}                       
    & \textbf{Android} & \textbf{709} & \textbf{2,579} 
    & \cmark & \cmark & \cmark & \cmark & \textbf{Coordinate} & \cmark \\
\bottomrule
\end{tabular}%
}
\vspace{-3mm}
\end{table*}

GUI agents---systems that perceive device interfaces and execute actions on behalf of users---have attracted growing research interest~\cite{hong2024cogagent, you2024ferret, cheng2024seeclick}. Powered by large foundational models, these agents interpret visual screens and perform operations such as tapping, swiping, and typing text, enabling task automation across smartphones~\cite{rawles2023androidinthewild}, desktops~\cite{xie2024osworld}, and web browsers~\cite{deng2023mind2web}.

A number of benchmarks have been developed to evaluate GUI agent capabilities (Table~\ref{tab:benchmark_comparison}). The majority of existing benchmarks provide only static screenshots as perceptual input. A few recent works have begun to incorporate additional modalities, introducing audio transcriptions~\cite{zheng2024gpt} or video recordings~\cite{chen2024gui, lin2024videogui, jang2024videowebarena}. Despite these advances, existing multimodal benchmarks largely treat audio and video as \emph{pre-task reference content}---for example, watching an instructional video before task execution. However, real-world device interaction routinely involves multimodal signals that are tightly coupled with the moment of action. On a typical smartphone, users encounter transient notification sounds, specific video playback states, or voice assistant instructions that directly govern the subsequent operation. These step-specific temporal and auditory contexts cannot be fully captured by static screenshots or pre-recorded reference videos. 

To address this gap, we introduce \textbf{OmniGUI} (Figure~\ref{fig:teaser}), the first benchmark designed to evaluate GUI agents receiving continuous, interleaved multimodal inputs---comprising static images, synchronous audio, and temporal video clips---at \emph{every} action step in real-world smartphone environments. OmniGUI encompasses 709 expert-demonstrated episodes (comprising 2,579 action steps) across 29 mobile applications. To ensure structural validity, the dataset is formulated around five cognitive operational dimensions (e.g., Temporal Reasoning, Instant Response) and subsequently categorized into three objective multimodal dependency levels (AV-Critical, AV-Supportive, AV-Present) based strictly on physical information availability. At each step, the agent is required to predict a precise action primitive and its corresponding parameters (e.g., normalized coordinates, strings) from a comprehensive 13-action space.

Our primary objective is to evaluate how GUI agents operate within fully multimodal interactive environments. Since dedicated omni-modal GUI agent frameworks are currently in their nascent stages, we select foundational omni-modal models (e.g., Gemini 3.0 Pro, Qwen3-Omni) capable of natively processing interleaved inputs to serve as agent proxies for our initial baselines. Furthermore, in the absence of official GUI-specific reasoning protocols for these models, we implement a standardized, deterministic inference pipeline utilizing a unified prompt template. This design ensures evaluation fairness and rigorously isolates the step-level perception-to-action capabilities. By establishing this standardized protocol, OmniGUI provides a reproducible foundation for assessing future purpose-built omni-modal agent architectures.

Our extensive evaluation across eight proprietary and open-source models reveals critical insights into the current state of multimodal action execution. The highest-performing model achieves an Exact Match (EM) step accuracy of 66.4\%, indicating that handling transient multimodal signals for precise step-level action prediction remains a significant challenge. Crucially, modality ablation studies empirically validate our dataset design: performance degrades significantly on AV-Critical tasks when non-visual modalities are removed, while remaining largely unaffected on purely static AV-Present tasks. Furthermore, the evaluation isolates specific operational bottlenecks in current architectures, such as cross-modal interference when presented with task-irrelevant multimodal signals, and significant performance degradation during concurrent dual-audio processing.

In summary, our contributions are as follows:
\begin{itemize}[leftmargin=*, nosep]
    \item We introduce OmniGUI, a GUI agent benchmark that provides interleaved image, audio, and video inputs at every action step, simulating the continuous multimodal perception required in real-world device interactions.
    \item We construct a high-quality, expert-demonstrated dataset of 709 episodes and 2,579 steps, systematically formulated around core HCI operational dimensions and rigorously annotated with objective multimodal dependency levels.
    \item We establish standardized initial baselines using foundational omni-modal models acting as agent proxies. Through comprehensive ablations, we validate the benchmark's structural necessity and identify specific operational bottlenecks (e.g., cross-modal interference) to provide empirical references for the development of future omni-agent frameworks.
\end{itemize}

\section{Related Work}

\subsection{GUI Agent Benchmarks}
\label{sec:related_gui}

The majority of existing GUI agent benchmarks rely exclusively on static screenshots as perceptual input. This includes extensive evaluations on Android~\cite{rawles2023androidinthewild, lu2025guiodyssey, rawles2024androidworld}, web browsers~\cite{deng2023mind2web}, desktop operating systems~\cite{xie2024osworld}, and cross-platform element grounding~\cite{cheng2024seeclick}. While these works have established the foundation for agentic automation~\cite{hong2024cogagent, you2024ferret, zhang2025appagent}, they fundamentally omit the auditory and temporal dynamics ubiquitous in real-world environments.

Recent efforts have begun incorporating non-visual modalities. Multimodal-Mind2Web~\cite{zheng2024gpt} augments web tasks with audio transcriptions, while GUI-World~\cite{chen2024gui} and VideoGUI~\cite{lin2024videogui} introduce video demonstrations for interaction analysis. Most related to our work is VideoWebArena~\cite{jang2024videowebarena}, which evaluates web agents using embedded multimedia content. However, these benchmarks predominantly treat audio and video as \emph{pre-task reference materials} rather than step-level synchronous inputs. OmniGUI diverges fundamentally by targeting mobile environments where transient multimodal signals (e.g., sound alerts, video playback states) are tightly coupled with the exact moment of action, requiring continuous perception-to-action grounding at every step.

\subsection{Omni-modal Foundation Models and Evaluations}
\label{sec:related_omni_perception}

The rapid evolution of foundational omni-modal models---capable of natively processing interleaved text, image, audio, and video---has been driven by both proprietary ecosystems (e.g., GPT-4o~\cite{hurst2024gpt}, Gemini family~\cite{team2024gemini, comanici2025gemini, gemini3report2025}) and open-source initiatives (e.g., Qwen3-Omni~\cite{xu2025qwen3omnitechnicalreport}, MiniCPM-o~\cite{yao2024minicpm}, VITA~\cite{fu2024vita}). 

Consequently, numerous benchmarks have been proposed to evaluate their multimodal capabilities. These include comprehensive tri-modal understanding evaluations~\cite{li2024omnibench, wang2025omnievalomnidirectionalautomaticrag}, multimodal conflict diagnostics~\cite{chowdhury2025avtrustbenchassessingenhancingreliability}, and broad audio-visual reasoning tasks~\cite{fu2025video, song2025video, yang2025audio, sakshi2024mmau}. Despite rigorous evaluation across diverse domains, these benchmarks share a critical limitation: they strictly assess \emph{passive perception and understanding}. The models output textual answers or classification labels based on fixed media inputs. None evaluate the sequential decision-making process where a model must translate dynamic, interleaved multimodal streams into executable operational primitives (e.g., coordinates, gestures) to alter the state of an interactive environment. OmniGUI bridges this exact gap, establishing a formal testbed for omni-modal agentic execution.

\section{The OmniGUI Benchmark}
\label{sec:benchmark}

\subsection{Interactive Environment and Formulation}
\label{sec:environment}

We formulate the mobile GUI interaction as a sequential decision-making process. At each step $t$, the omni-modal agent receives a comprehensive observation state $S_t$ from the environment and predicts an executable action $a_t$ to fulfill a given natural language instruction $G$. 

The observation state $S_t$ is defined as a tuple of multimodal inputs: $S_t = (I_t, V_t, A_t, H_t)$, where:
\begin{itemize}[leftmargin=*, nosep]
    \item $I_t$ is the high-resolution static screenshot captured at the current step $t$.
    \item $V_t$ is the temporal video clip recording the screen dynamics from the previous action execution up to step $t$.
    \item $A_t$ is the synchronous audio stream corresponding to $V_t$, capturing system sounds, media playback, or user voice commands.
    \item $H_t = \{a_1, a_2, \dots, a_{t-1}\}$ represents the historical action trajectory.
\end{itemize}

Based on the instruction $G$ and the multimodal state $S_t$, the agent generates an action $a_t \in \mathcal{A}$. As detailed in Table~\ref{tab:action_space}, the action space $\mathcal{A}$ encompasses 13 operational primitives across five categories: wait/observe (\texttt{NONE}), positional actions (e.g., \texttt{TAP}), gestural actions (e.g., \texttt{SWIPE\_UP}), text input (\texttt{INPUT}), and system/status signals (e.g., \texttt{HOME}, \texttt{TASK\_COMPLETE}). Continuous coordinate parameters $(x, y)$ are normalized to a resolution-independent $[0, 1000] \times[0, 1000]$ scale.

\begin{table}[t]
\centering
\captionsetup{skip=3pt}
\caption{
\textbf{Definition of the OmniGUI Action Space.} 
The agent predicts an action primitive along with its parameters. Coordinates $(x,y)$ are normalized to the resolution-independent range $[0, 1000]^2$. 
}
\label{tab:action_space}
\vspace{1mm}
\footnotesize 
\renewcommand{\arraystretch}{0.95} 
\setlength{\tabcolsep}{4pt} 

\begin{tabularx}{\linewidth}{@{} l l >{\raggedright\arraybackslash}X @{}}
\toprule
\textbf{Action Primitive} & \textbf{Parameter Space} & \textbf{Semantics / Usage} \\
\midrule

\rowcolor{gray!10} \multicolumn{3}{l}{\textit{Temporal / Idle}} \\
\texttt{NONE} (-1)     & $\varnothing$           & Wait or observe without taking action \\

\midrule
\rowcolor{gray!10} \multicolumn{3}{l}{\textit{Positional Actions}} \\
\texttt{TAP} (0)       & $(x,y) \in [0, 1000]^2$ & Click a UI element (button, link, icon) \\
\texttt{DOUBLE\_TAP} (1)& $(x,y) \in[0, 1000]^2$ & Trigger specific states (e.g., zoom, like) \\
\texttt{LONG\_PRESS} (2)& $(x,y) \in[0, 1000]^2$ & Open context menus or select items \\

\midrule
\rowcolor{gray!10} \multicolumn{3}{l}{\textit{Gestural Actions}} \\
\texttt{SWIPE\_UP} (3) & $(x,y) \in [0, 1000]^2$ & Scroll down content or feed \\
\texttt{SWIPE\_DOWN} (4)& $(x,y) \in[0, 1000]^2$ & Refresh page or scroll up \\
\texttt{SWIPE\_LEFT} (5)& $(x,y) \in [0, 1000]^2$ & Navigate carousels or switch tabs \\
\texttt{SWIPE\_RIGHT} (6)& $(x,y) \in[0, 1000]^2$ & Navigate back or switch tabs \\

\midrule
\rowcolor{gray!10} \multicolumn{3}{l}{\textit{Text Input}} \\
\texttt{INPUT} (7)     & $\mathcal{S}$ (String)  & Enter text into a focused field \\

\midrule
\rowcolor{gray!10} \multicolumn{3}{l}{\textit{System \& Status Signals}} \\
\texttt{BACK} (8)      & $\varnothing$           & Return to previous screen/activity \\
\texttt{HOME} (9)      & $\varnothing$           & Return to device home screen \\
\texttt{TASK\_COMPLETE} (10)& $\varnothing$      & Signal successful task completion \\
\texttt{TASK\_IMPOSSIBLE} (11)& $\varnothing$    & Signal task is infeasible/stuck \\

\bottomrule
\end{tabularx}
\vspace{-5mm}
\end{table}

\subsection{Task Taxonomy and Dataset Statistics}
\label{sec:task_taxonomy}

\begin{figure*}[t]
\centering
\includegraphics[width=\textwidth]{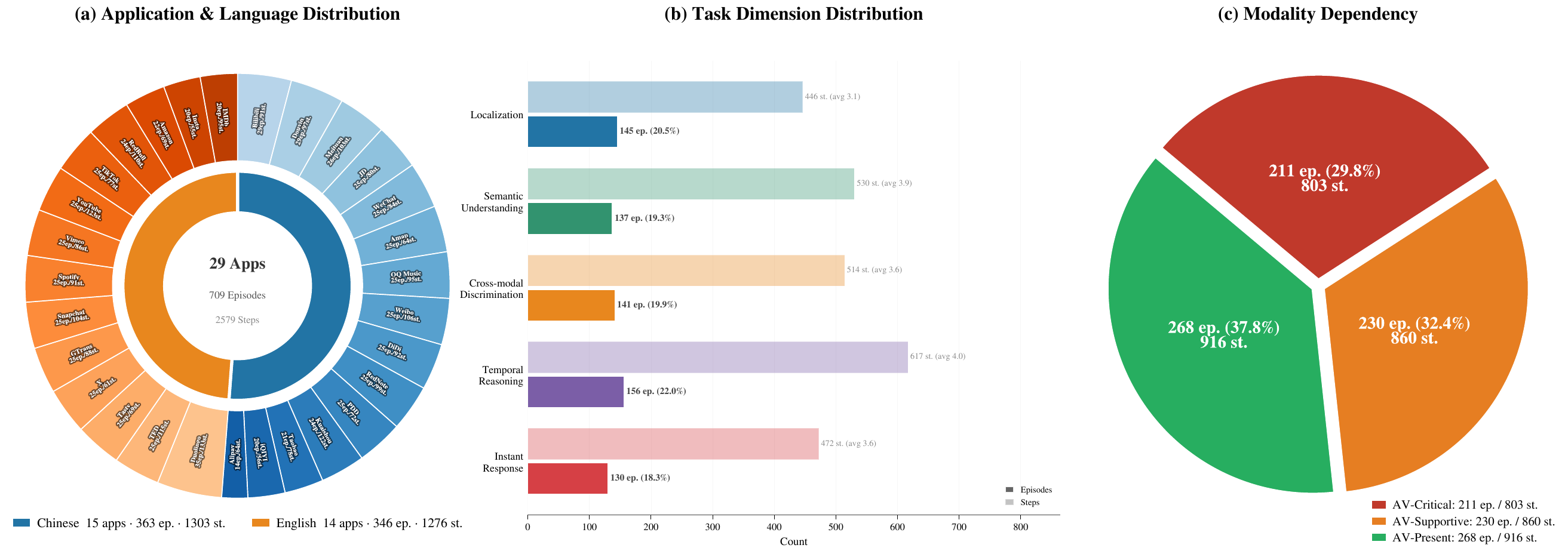}
\caption{
\textbf{Dataset statistics of OmniGUI.} 
(a) Application and language distribution, detailing the composition of 709 episodes and 2,579 fine-grained steps across 29 smartphone applications. 
(b) Distribution of episodes and steps across five core task dimensions, which are grounded in human-computer interaction and multimodal cognitive processes. 
(c) Proportion of episodes and steps categorized by multimodal dependency levels, derived objectively from GUI information availability.
}
\label{fig:dataset_overview}
\end{figure*}

The OmniGUI benchmark comprises 709 multi-step episodes, yielding a total of 2,579 fine-grained action steps (averaging 3.64 steps per episode). Constructed across 29 widely used smartphone applications, the dataset maintains a balanced bilingual distribution to assess cross-lingual generalization, including 15 Chinese applications (363 episodes, 1,303 steps) and 14 English applications (346 episodes, 1,276 steps), as illustrated in Figure~\ref{fig:dataset_overview}(a). We organize the benchmark along two primary analytical axes: \emph{task dimension} and \emph{multimodal dependency}.

\paragraph{Task Dimensions and Formulation.}
To systematically evaluate the capabilities of omni-modal GUI agents, we established a top-down task taxonomy drawing upon Human-Computer Interaction (HCI) principles. We defined five operational dimensions that map the cognitive processing flow required for agentic execution---spanning perception, comprehension, reasoning, and reaction:
\begin{itemize}[leftmargin=*, nosep]
    \item \textbf{Localization (20.5\% ep. / 446 steps):} Grounding actions to specific spatial coordinates based on visual or auditory descriptions.
    \item \textbf{Semantic Understanding (19.3\% ep. / 530 steps):} Comprehending textual, visual, or spoken semantics to formulate multi-step execution plans.
    \item \textbf{Cross-modal Discrimination (19.9\% ep. / 514 steps):} Synthesizing and aligning complementary information across video, audio, and text modalities.
    \item \textbf{Temporal Reasoning (22.0\% ep. / 617 steps):} Tracking dynamic UI changes, moving elements, or event sequences over time.
    \item \textbf{Instant Response (18.3\% ep. / 472 steps):} Reacting promptly to transient auditory or visual cues, such as alarms or specific video frames.
\end{itemize}
Guided by these five predefined dimensions, our annotators formulated the 709 goal-oriented episodes across 29 applications. This top-down formulation ensures that the collected tasks are not only ecologically authentic but also provide balanced coverage across different cognitive complexities.

\paragraph{Multimodal Dependency Taxonomy.}
To systematically quantify how omni-modal agents utilize non-visual signals, we categorize all episodes into three dependency levels (Figure~\ref{fig:dataset_overview}c). This categorization is based solely on the objective information structure of the GUI environment (i.e., the physical availability of task-relevant signals) and is independent of empirical model performance. We define the following annotation codebook:

\begin{itemize}[leftmargin=*, topsep=2pt, itemsep=4pt]
    \item \textbf{AV-Critical (29.8\% ep. / 803 steps):} The correct action for at least one step cannot be determined from the static screenshot alone. The decision-critical information is exclusively present in the audio stream (e.g., a spoken instruction, a specific ringtone) or the temporal video stream (e.g., timing an action to a specific playback state).
    
    \item \textbf{AV-Supportive (32.4\% ep. / 860 steps):} The static screenshot contains sufficient information to deduce the next action, but audio or video provides corroborating context that reduces ambiguity (e.g., background audio confirming an active media state). Non-visual signals improve robustness but are not strictly mandatory.
    
    \item \textbf{AV-Present (37.8\% ep. / 916 steps):} Purely static UI tasks where all steps are fully resolvable from the static screenshot and action history. Audio and video modalities are present as environmental background noise and carry no additional task-relevant information.
\end{itemize}

\paragraph{Annotation Procedure and Quality Assurance.}
Following the task collection, we conducted a post-hoc evaluation to assign the multimodal dependency labels to each episode. To implement this, we established a strict modality-ablated annotation procedure. For each step, annotators were initially provided with only the static screenshot to determine if the correct action was unambiguously resolvable. Subsequently, the temporal video and audio streams were revealed, allowing them to finalize the objective dependency level based on whether the non-visual modalities introduced essential information.

To quantify the reliability of this taxonomy, a random subset of 100 episodes was independently annotated by a second reviewer. The process yielded a high inter-annotator agreement (Cohen's $\kappa = 0.84$), confirming substantial objective consensus. Disagreements in edge cases were resolved by a third senior annotator via majority vote.

\subsection{Data Collection and Annotation Pipeline}
\label{sec:data_collection}

The construction of OmniGUI follows a systematic pipeline designed to elicit diverse and high-quality human demonstrations. 

\paragraph{Task Formulation and Annotator Demographics.}
To operationalize the top-down taxonomy established in Section~\ref{sec:task_taxonomy}, we recruited 10 native smartphone users, each with over five years of daily Android operating experience. Guided strictly by the five predefined cognitive dimensions, these experienced annotators ideated and formulated goal-oriented usage scenarios across 29 diverse applications. This protocol ensures that the dataset achieves systematic theoretical coverage while maintaining authentic ecological validity.

\paragraph{Demonstration Recording.}
For each formulated task, the expert annotators executed the intended trajectory on physical Android devices. A background logging system synchronously captured the screen video at 30 frames per second (FPS), the internal device audio, and the precise touch interaction events. These 709 recorded human demonstrations serve as the optimal ground-truth trajectories for our evaluation. Screenshots $I_t$ were extracted at the exact timestamp preceding each human action $a_t$. The video clip $V_t$ and audio segment $A_t$ for each step were segmented using the interval between the completion of $a_{t-1}$ and the initiation of $a_t$.

\paragraph{Formalized Annotation.}
We developed a dedicated web-based annotation platform for multimodal GUI tasks. Annotators utilized this platform to transcribe the raw touch events into the formalized action space $\mathcal{A}$. For positional and gestural actions, annotators verified the target UI elements and bounded the normalized coordinates. For text inputs, the exact alphanumeric strings were recorded. Finally, each episode was assigned its objective multimodal dependency label as defined in Section~\ref{sec:task_taxonomy}.

\subsection{Evaluation Protocol and Metrics}
\label{sec:evaluation}

We evaluate the models using a step-level teacher-forcing protocol, which isolates per-step multimodal perception capabilities from cascading compounding errors typical in autonomous rollouts. At each step $t$, the model receives the ground-truth history $H_t$ and predicts $a_t$. Because our dataset is built upon expert human demonstrations, achieving 100\% performance conceptually represents perfect alignment with expert human operational intent. We employ four quantitative metrics:

\begin{itemize}[leftmargin=*, nosep]
    \item \textbf{Type Match (TM) [Step-level]:} Calculates the accuracy of predicting the correct action primitive (e.g., selecting \texttt{TAP} instead of \texttt{SWIPE\_UP}), disregarding the specific parameters.
    \item \textbf{Exact Match (EM) [Step-level]:} A step is considered an exact match if both the action primitive and its associated parameters are correct. For positional actions, the predicted coordinates $(x, y)$ must fall within the bounding box of the ground-truth target UI element. For text inputs, the generated string must exactly match the target text.
    \item \textbf{Success Rate (SR)[Episode-level]:} An episode is marked successful ($1.0$) if and only if the EM condition is satisfied for every single step within the trajectory; otherwise, it is $0.0$.
    \item \textbf{Goal Progress (GP) [Episode-level]:} Measures the partial completion rate of a multi-step episode. It is calculated as the ratio of correctly executed steps (EM) to the total number of steps within that specific episode's ground-truth trajectory. This provides a granular, step-aware assessment for complex tasks even when the overall episode ultimately fails.
\end{itemize}

\section{Experiments}
\label{sec:experiments}

This section presents the experimental evaluation of OmniGUI. The experiments are structured to achieve two primary objectives: first, to empirically validate the structural design and necessity of the proposed multimodal benchmark mechanisms; and second, to establish initial performance baselines for omni-modal GUI agents. Because dedicated omni-agent frameworks are currently in their nascent stage, we utilize foundational omni-modal models as direct proxies to execute the interactive tasks. We outline the experimental setup (Section~\ref{sec:experimental_setup}), present the overall evaluation results (Section~\ref{sec:main_results}), conduct modality ablation analyses to verify our task taxonomy (Section~\ref{sec:ablation}), and conclude with a qualitative error analysis (Section~\ref{sec:error_analysis}).

\subsection{Experimental Setup}
\label{sec:experimental_setup}

\paragraph{Evaluated Models.}
We evaluate state-of-the-art proprietary models: Gemini 3.0 Pro~\cite{gemini3report2025}, Gemini 3.0 Flash~\cite{gemini3report2025}, Gemini 2.5 Pro~\cite{comanici2025gemini}, and Gemini 2.5 Flash~\cite{comanici2025gemini}. We also evaluate leading open-source models: Qwen3-Omni~\cite{xu2025qwen3omnitechnicalreport}, MiniCPM-o 4.5~\cite{yao2024minicpm}, VITA-1.5~\cite{fu2025vita}, and Baichuan-Omni-1.5~\cite{li2025baichuan}.\footnote{GPT-4o is excluded from the current evaluation. Its Chat Completions API lacks native support for interleaved raw audio-visual ingestion, while the Realtime API operates as a low-latency speech-to-speech stream, which is incompatible with the deterministic, step-level multimodal batch evaluation required by our benchmark protocol.}

\paragraph{Prompt Design and Input Structure.}
To evaluate perception-to-action capabilities without agent-specific prompt engineering, we adopt a unified prompt consisting of a system instruction and a user message. The system prompt defines the Android GUI agent persona, specifies the complete action space (11 action primitives plus a wait/observe option), establishes the normalized $[0, 1000] \times [0, 1000]$ coordinate system, and strictly enforces a single JSON object as the output format. The user message structures the step-level context using an interleaved multimodal sequence. It sequentially presents the historical screenshot from step $t{-}2$ (if available), the current-step video clip, the synchronous environment audio, the current static screenshot, and the text-based task goal. To maintain ecological validity, the textual task instruction is adaptively provided in either Chinese or English, matching the native language of the target application. The ground-truth action history is provided as a structured text list of previously executed action types and parameters. The exact prompt templates and raw JSON data examples are provided in the supplementary material.

\paragraph{Implementation Details.}
Model-specific adaptations are strictly limited to API-level payload formatting. To minimize sampling variance and obtain the models' most confident decision boundaries, we employ deterministic greedy decoding by setting the generation temperature to $0.0$ and \texttt{do\_sample = False} across all frameworks where explicit parameter control is supported. A maximum generation limit of 4096 tokens is applied. Comprehensive hardware configurations and exact API settings are detailed in the supplementary material.

\subsection{Main Results}
\label{sec:main_results}

Table~\ref{tab:main_comprehensive} presents the overall performance metrics and a fine-grained dimension breakdown for all evaluated models. 

Among the proprietary models, Gemini 3.0 Pro achieves the highest overall performance, yielding an Exact Match (EM) of 66.4\% and a Success Rate (SR) of 33.1\%. Despite being the state-of-the-art, its absolute success rate remains low, indicating that executing multi-step GUI tasks with interleaved transient multimodal signals remains a significant bottleneck for current models. Gemini 3.0 Flash follows closely, occasionally surpassing the Pro version in specific dimensions such as Temporal Reasoning.

The evaluation reveals a substantial capability gap between proprietary and open-source models. Qwen3-Omni leads the open-source category with an EM of 33.4\% and an SR of 5.2\%, while the remaining open-source models struggle to complete full episodes successfully (SR $\le$ 1.1\%). 

Across the five cognitive dimensions, performance varies consistently. Models generally exhibit higher Exact Match scores on static Localization tasks (e.g., 79.9\% for Gemini 3.0 Pro) compared to Cross-modal Discrimination (59.9\%) or Temporal Reasoning (61.8\%) tasks. This variation objectively reflects the increased complexity of integrating dynamic temporal and auditory cues into precise spatial actions compared to traditional screenshot-only visual grounding.

\begin{table*}[t]
\centering
\caption{
\textbf{Comprehensive evaluation results on OmniGUI.}
We report performance across four metrics: Type Match (TM), Exact Match (EM), Success Rate (SR), and Goal Progress (GP). The table presents both the \textbf{Overall} performance and a fine-grained breakdown across five specific task dimensions: Localization (\textit{Local.}), Semantic Understanding (\textit{Semantic Understand.}), Cross-modal Discrimination (\textit{Cross-modal Discr.}), Temporal Reasoning (\textit{Temporal Reason.}), and Instant Response (\textit{Instant Resp.}). All metrics are reported as percentages (\%). \textbf{Bold} indicates the best performance in each respective column.
}
\label{tab:main_comprehensive}
\vspace{-2mm}

\setlength{\tabcolsep}{1.5pt}      
\renewcommand{\arraystretch}{1.35} 

\resizebox{\textwidth}{!}{%
\begin{tabular}{@{} l cccc cccc cccc cccc cccc cccc @{}}
\toprule
\multirow{3.5}{*}{\textbf{Model}}
& \multicolumn{4}{c}{\cellcolor{gray!20}\textbf{Overall}}
& \multicolumn{4}{c}{\makecell{\textbf{Localization}}}
& \multicolumn{4}{c}{\makecell{\textbf{Semantic}\\\textbf{Understand.}}}
& \multicolumn{4}{c}{\makecell{\textbf{Cross-modal}\\\textbf{Discr.}}}
& \multicolumn{4}{c}{\makecell{\textbf{Temporal}\\\textbf{Reason.}}}
& \multicolumn{4}{c}{\makecell{\textbf{Instant}\\\textbf{Resp.}}} \\

\arrayrulecolor{gray!20}\cmidrule{2-5}\arrayrulecolor{black}\cmidrule(lr){6-9} \cmidrule(lr){10-13} \cmidrule(lr){14-17} \cmidrule(lr){18-21} \cmidrule(l){22-25}

& \cellcolor{gray!20}\textbf{TM} & \cellcolor{gray!20}\textbf{EM} & \cellcolor{gray!20}\textbf{SR} & \cellcolor{gray!20}\textbf{GP}
& \textbf{TM} & \textbf{EM} & \textbf{SR} & \textbf{GP}
& \textbf{TM} & \textbf{EM} & \textbf{SR} & \textbf{GP}
& \textbf{TM} & \textbf{EM} & \textbf{SR} & \textbf{GP}
& \textbf{TM} & \textbf{EM} & \textbf{SR} & \textbf{GP}
& \textbf{TM} & \textbf{EM} & \textbf{SR} & \textbf{GP} \\
\midrule
\rowcolor{gray!10} \multicolumn{25}{l}{\textit{Proprietary Models}} \\
\midrule
Gemini 3 Pro~\cite{gemini3report2025}
& \cellcolor{gray!20}\textbf{80.0} & \cellcolor{gray!20}\textbf{63.6} & \cellcolor{gray!20}\textbf{33.4} & \cellcolor{gray!20}\textbf{43.6}
& \textbf{86.3} & \textbf{76.2} & \textbf{55.9} & 62.6
& \textbf{77.4} & \textbf{61.1} & \textbf{31.4} & \textbf{42.0}
& \textbf{76.6} & \textbf{59.1} & \textbf{30.1} & \textbf{41.3}
& 78.9 & \textbf{61.0} & 22.7 & 36.9
& \textbf{81.8} & \textbf{62.7} & \textbf{27.6} & \textbf{35.6} \\
Gemini 3 Flash~\cite{gemini3report2025}
& \cellcolor{gray!20}78.3 & \cellcolor{gray!20}61.3 & \cellcolor{gray!20}30.3 & \cellcolor{gray!20}43.5
& 85.0 & 75.6 & 53.1 & \textbf{63.1}
& 75.3 & 58.5 & 25.5 & 41.1
& 72.8 & 56.0 & 23.5 & 38.7
& \textbf{80.0} & 60.3 & \textbf{25.3} & \textbf{39.4}
& 79.2 & 57.9 & 22.8 & 34.2 \\
Gemini 2.5 Pro~\cite{comanici2025gemini}
& \cellcolor{gray!20}75.7 & \cellcolor{gray!20}44.1 & \cellcolor{gray!20}15.5 & \cellcolor{gray!20}26.3
& 86.1 & 58.1 & 31.7 & 41.5
& 72.8 & 37.7 & 11.7 & 22.4
& 70.6 & 40.1 & 13.2 & 25.1
& 73.8 & 44.3 & 9.7 & 22.5
& 76.6 & 42.1 & 11.0 & 19.5 \\
Gemini 2.5 Flash~\cite{comanici2025gemini}
& \cellcolor{gray!20}69.5 & \cellcolor{gray!20}37.8 & \cellcolor{gray!20}12.4 & \cellcolor{gray!20}24.5
& 75.1 & 50.9 & 29.0 & 42.6
& 70.4 & 34.3 & 8.0 & 18.2
& 64.9 & 35.7 & 11.8 & 25.3
& 67.7 & 35.1 & 9.1 & 21.8
& 71.0 & 34.5 & 3.9 & 13.7 \\
\midrule
\rowcolor{gray!10} \multicolumn{25}{l}{\textit{Open-source Models}} \\
\midrule
Qwen3-Omni~\cite{xu2025qwen3omnitechnicalreport}
& \cellcolor{gray!20}63.1 & \cellcolor{gray!20}32.3 & \cellcolor{gray!20}5.1 & \cellcolor{gray!20}17.4
& 65.7 & 42.4 & 10.3 & 28.5
& 58.3 & 29.6 & 2.9 & 14.0
& 57.9 & 26.2 & 2.2 & 13.2
& 66.2 & 31.1 & 5.8 & 16.8
& 67.4 & 33.7 & 3.9 & 13.7 \\
VITA-1.5~\cite{fu2025vita}
& \cellcolor{gray!20}39.3 & \cellcolor{gray!20}12.1 & \cellcolor{gray!20}1.1 & \cellcolor{gray!20}2.2
& 48.4 & 14.8 & 2.8 & 3.9
& 43.4 & 16.4 & 2.2 & 3.2
& 33.9 & 11.5 & 0.0 & 0.8
& 35.4 & 7.7 & 0.6 & 2.0
& 36.9 & 10.3 & 0.0 & 0.8 \\
MiniCPM-o-4.5~\cite{yao2024minicpm}
& \cellcolor{gray!20}32.8 & \cellcolor{gray!20}4.8 & \cellcolor{gray!20}0.1 & \cellcolor{gray!20}1.4
& 34.8 & 7.4 & 0.7 & 2.2
& 34.7 & 5.5 & 0.0 & 1.0
& 25.2 & 4.4 & 0.0 & 2.2
& 34.8 & 3.9 & 0.0 & 0.6
& 33.3 & 3.2 & 0.0 & 0.8 \\
Baichuan-Omni-1.5~\cite{li2025baichuan}
& \cellcolor{gray!20}17.0 & \cellcolor{gray!20}3.3 & \cellcolor{gray!20}0.0 & \cellcolor{gray!20}0.4
& 19.5 & 4.9 & 0.0 & 1.0
& 16.2 & 4.0 & 0.0 & 0.5
& 12.9 & 1.4 & 0.0 & 0.0
& 18.2 & 2.3 & 0.0 & 0.2
& 18.2 & 4.1 & 0.0 & 0.5 \\
\bottomrule
\end{tabular}%
}
\end{table*}

\subsection{Ablation Analysis}
\label{sec:ablation}

To empirically validate our task taxonomy and observe how models utilize different modalities, we conduct two sets of ablation studies using representative proprietary (Gemini 3 Pro, Gemini 2.5 Flash) and open-source (Qwen3-Omni) models.

\paragraph{Modality Ablation Analysis.}
Table~\ref{tab:modality_ablation_full} presents the results of systematically masking audio and video inputs. The observed performance degradation strictly aligns with our human-annotated multimodal dependency levels, verifying the structural validity of the OmniGUI benchmark. For instance, completely removing audio and video inputs (\textit{No AV}) causes the most severe performance drop on AV-Critical tasks across all models (e.g., a 10.5\% Exact Match drop for Gemini 3 Pro). Conversely, on purely static AV-Present tasks, removing these modalities yields negligible performance variation ($-0.3\%$). 

Furthermore, the ablation results expose a cross-modal interference phenomenon. For Gemini 2.5 Flash and Qwen3-Omni, providing the full multimodal input (I+A+V) on AV-Present tasks results in lower performance compared to providing the static image alone (\textit{No AV}). Specifically, Gemini 2.5 Flash's EM score decreases from 49.9\% to 40.8\% when environmental audio and video are introduced. This empirically indicates that the inclusion of task-irrelevant multimodal signals can negatively impact action prediction accuracy in visually sufficient contexts.

\paragraph{Instruction Modality (Text vs. TTS).}
In realistic agent deployments, users often initiate tasks via spoken commands. Table~\ref{tab:tts_ablation_full} compares model performance when text instructions are replaced with Text-to-Speech (TTS) synthesized audio, while keeping all environmental multimodal inputs intact. 

The evaluation reveals an asymmetric performance degradation. On static AV-Present tasks, processing a spoken instruction incurs virtually no penalty (e.g., $\Delta \approx 0.1\%$ EM for Gemini 3 Pro). However, on AV-Critical tasks, substituting text with TTS causes a uniform and pronounced drop across the evaluated models ($-5.3\%$ EM for Gemini 3 Pro). This contrast isolates a specific difficulty in concurrent multimodal processing: while current models maintain performance when grounding a single spoken instruction to a static image, they exhibit significant degradation when required to process the spoken instruction simultaneously with environmental audio cues and dynamic video frames.

\begin{table*}[t]
\centering
\caption{
\textbf{Modality ablation across dependency levels.} We report all metrics (\%) and show the performance gap inline ($\Delta = \text{Ablated} - \text{Full}$). To highlight the most impactful modalities, for each model and metric (column), the most severe performance drop among the three ablation settings is marked in \textbf{\textcolor{red}{bold red}}. Conversely, the most significant anomalous performance gain is marked in \textbf{\textcolor{teal}{bold teal}}, revealing strong cross-modal interference when irrelevant modalities are provided in Present tasks.
}
\label{tab:modality_ablation_full}
\vspace{-2mm}

\newcommand{\cD}[2]{\makecell[c]{#1 \\[-2pt] \textcolor{red!70!black}{\fontsize{7pt}{8pt}\selectfont(#2)}}}     
\newcommand{\cG}[2]{\makecell[c]{#1 \\[-2pt] \textcolor{teal!90!black}{\fontsize{7pt}{8pt}\selectfont(+#2)}}}    
\newcommand{\cBD}[2]{\makecell[c]{#1 \\[-2pt] \textbf{\textcolor{red}{\fontsize{7pt}{8pt}\selectfont(#2)}}}}    
\newcommand{\cBG}[2]{\makecell[c]{#1 \\[-2pt] \textbf{\textcolor{teal}{\fontsize{7pt}{8pt}\selectfont(+#2)}}}}   

\setlength{\tabcolsep}{2.5pt} 
\renewcommand{\arraystretch}{1.45} 

\resizebox{\textwidth}{!}{%
\begin{tabular}{@{} l l cccc @{\hspace{5pt}} cccc @{\hspace{5pt}} cccc @{\hspace{5pt}} cccc @{}}
\toprule
\multirow{2.5}{*}{\textbf{Model}} & \multirow{2.5}{*}{\textbf{Modality Input}} 
& \multicolumn{4}{c@{\hspace{5pt}}}{\makecell{\textbf{AV-Critical} \\ \textbf{(34.9\%)}}} 
& \multicolumn{4}{c@{\hspace{5pt}}}{\makecell{\textbf{AV-Supportive} \\ \textbf{(38.6\%)}}} 
& \multicolumn{4}{c@{\hspace{5pt}}}{\makecell{\textbf{AV-Present} \\ \textbf{(26.5\%)}}} 
& \multicolumn{4}{c}{\makecell{\textbf{Overall} \\ \textbf{(100\%)}}} \\
\cmidrule(lr){3-6} \cmidrule(lr){7-10} \cmidrule(lr){11-14} \cmidrule(l){15-18}
& & \textbf{TM} & \textbf{EM} & \textbf{SR} & \textbf{GP} 
  & \textbf{TM} & \textbf{EM} & \textbf{SR} & \textbf{GP} 
  & \textbf{TM} & \textbf{EM} & \textbf{SR} & \textbf{GP} 
  & \textbf{TM} & \textbf{EM} & \textbf{SR} & \textbf{GP} \\
\midrule

\rowcolor{highlightgray!50} \multicolumn{18}{l}{\textit{Proprietary Models}} \\
\midrule

\multirow{4}{*}{Gemini 3 Pro} 
& Full (I+A+V) & 76.9 & 57.9 & 33.2 & 42.2 & 79.6 & 65.2 & 34.4 & 45.0 & 84.7 & 69.0 & 33.0 & 44.4 & 80.0 & 63.6 & 33.4 & 43.6 \\
& No Audio (I+V) & \cD{74.4}{-2.5} & \cD{55.9}{-2.0} & \cD{28.7}{-4.5} & \cD{39.6}{-2.6} & 79.6 & \cD{64.1}{-1.1} & \cD{34.1}{-0.3} & \cD{44.5}{-0.5} & \cBG{85.0}{0.3} & \cBG{70.9}{1.9} & \cG{36.2}{3.2} & \cG{46.9}{2.5} & \cD{79.2}{-0.8} & \cD{63.0}{-0.6} & \cD{32.7}{-0.7} & \cD{43.3}{-0.3} \\
& No Video (I+A) & \cD{69.1}{-7.8} & \cD{50.0}{-7.9} & \cBD{16.0}{-17.2} & \cBD{34.5}{-7.7} & \cBD{74.6}{-5.0} & \cD{59.2}{-6.0} & \cBD{25.2}{-9.2} & \cD{41.5}{-3.5} & \cBD{81.3}{-3.4} & \cBD{67.7}{-1.3} & \cG{35.1}{2.1} & \cG{48.0}{3.6} & \cD{74.4}{-5.6} & \cD{58.1}{-5.5} & \cBD{24.4}{-9.0} & \cBD{40.7}{-2.9} \\
& No AV (Img Only) & \cBD{67.3}{-9.6} & \cBD{48.9}{-9.0} & \cD{17.2}{-16.0} & \cD{34.5}{-7.7} & \cD{75.3}{-4.3} & \cBD{59.0}{-6.2} & \cD{26.3}{-8.1} & \cBD{41.0}{-4.0} & \cD{81.8}{-2.9} & \cD{68.9}{-0.1} & \cBG{37.8}{4.8} & \cBG{48.8}{4.4} & \cBD{74.2}{-5.8} & \cBD{58.0}{-5.6} & \cD{26.1}{-7.3} & \cD{40.8}{-2.8} \\

\midrule

\multirow{4}{*}{Gem. 2.5 Flash} 
& Full (I+A+V) & 66.1 & 35.4 & 13.9 & 26.3 & 70.0 & 38.7 & 14.1 & 25.9 & 73.9 & 39.3 & 8.6 & 20.5 & 69.5 & 37.8 & 12.4 & 24.5 \\
& No Audio (I+V) & \cD{61.6}{-4.5} & \cD{32.7}{-2.7} & \cD{9.0}{-4.9} & \cD{22.1}{-4.2} & \cD{68.8}{-1.2} & \cD{36.7}{-2.0} & \cD{11.5}{-2.6} & \cD{23.6}{-2.3} & \cBG{76.0}{2.1} & \cG{41.7}{2.4} & \cG{10.3}{1.7} & \cG{23.2}{2.7} & \cD{68.2}{-1.3} & \cD{36.8}{-1.0} & \cD{10.3}{-2.1} & \cD{23.0}{-1.5} \\
& No Video (I+A) & \cD{60.7}{-5.4} & \cBD{26.4}{-9.0} & \cBD{2.0}{-11.9} & \cBD{14.9}{-11.4} & \cD{66.1}{-3.9} & \cBD{31.9}{-6.8} & \cBD{5.9}{-8.2} & \cBD{17.5}{-8.4} & \cBD{69.7}{-4.2} & \cBD{37.2}{-2.1} & \cBD{7.6}{-1.0} & \cBD{20.3}{-0.2} & \cD{65.1}{-4.4} & \cBD{31.4}{-6.4} & \cBD{4.9}{-7.5} & \cBD{17.3}{-7.2} \\
& No AV (Img Only) & \cBD{57.2}{-8.9} & \cD{28.2}{-7.2} & \cD{4.1}{-9.8} & \cD{18.1}{-8.2} & \cBD{65.4}{-4.6} & \cD{36.2}{-2.5} & \cD{8.5}{-5.6} & \cD{23.6}{-2.3} & \cD{72.1}{-1.8} & \cBG{48.3}{9.0} & \cBG{15.7}{7.1} & \cBG{34.4}{13.9} & \cBD{64.2}{-5.3} & \cD{36.5}{-1.3} & \cD{8.7}{-3.7} & \cD{24.4}{-0.1} \\

\midrule
\rowcolor{highlightgray!50} \multicolumn{18}{l}{\textit{Open-source Model}} \\
\midrule

\multirow{4}{*}{Qwen3-Omni} 
& Full (I+A+V) & 58.0 & 29.4 & 7.0 & 18.5 & 65.1 & 33.5 & 4.8 & 17.4 & 66.9 & 34.4 & 3.2 & 16.0 & 63.1 & 32.3 & 5.1 & 17.4 \\
& No Audio (I+V) & \cD{57.9}{-0.1} & \cD{26.7}{-2.7} & \cD{5.7}{-1.3} & \cD{17.3}{-1.2} & \cD{64.0}{-1.1} & \cD{32.7}{-0.8} & \cBD{4.4}{-0.4} & \cBG{17.9}{0.5} & \cBG{67.2}{0.3} & \cG{34.9}{0.5} & \cG{3.2}{0.0} & \cG{16.7}{0.7} & \cD{62.7}{-0.4} & \cBD{31.1}{-1.2} & \cBD{4.5}{-0.6} & \cBD{17.3}{-0.1} \\
& No Video (I+A) & \cD{57.1}{-0.9} & \cD{28.8}{-0.6} & \cD{4.5}{-2.5} & \cD{17.8}{-0.7} & \cD{62.6}{-2.5} & \cBD{31.2}{-2.3} & \cBG{5.6}{0.8} & \cBD{16.3}{-1.1} & \cBD{66.0}{-0.9} & \cBG{39.3}{4.9} & \cBG{7.0}{3.8} & \cG{20.1}{4.1} & \cD{61.6}{-1.5} & \cBG{32.6}{0.3} & \cBG{5.6}{0.5} & \cBG{17.9}{0.5} \\
& No AV (Img Only) & \cBD{54.7}{-3.3} & \cBD{25.6}{-3.8} & \cBD{3.7}{-3.3} & \cBD{15.6}{-2.9} & \cBD{62.3}{-2.8} & \cD{31.4}{-2.1} & \cG{4.8}{0.0} & \cD{16.9}{-0.5} & \cD{66.2}{-0.7} & \cG{38.5}{4.1} & \cG{7.0}{3.8} & \cBG{21.8}{5.8} & \cBD{60.6}{-2.5} & \cD{31.3}{-1.0} & \cD{4.9}{-0.2} & \cG{17.7}{0.3} \\

\bottomrule
\end{tabular}%
}
\vspace{-3mm}
\end{table*}

\begin{table*}[t]
\centering
\caption{
\textbf{Impact of instruction modality (Text vs. TTS Voice).} We report performance metrics (\%) and show the gap inline ($\Delta = \text{TTS} - \text{Text}$). To objectively highlight the cognitive load distribution, for each model and metric (across rows), the most severe performance drop among the three dependency levels is marked in \textbf{\textcolor{red}{bold red}}. Conversely, any performance maintenance or gain ($\Delta \ge 0$) is marked in \textbf{\textcolor{teal}{bold teal}}. This reveals that dual-audio stream processing (TTS + environmental audio) uniformly degrades performance on multimodal-dependent tasks, while static AV-Present tasks remain largely immune.
}
\label{tab:tts_ablation_full}
\vspace{-2mm}

\newcommand{\cD}[2]{\makecell[c]{#1 \\[-2pt] \textcolor{red!70!black}{\fontsize{7pt}{8pt}\selectfont(#2)}}}
\newcommand{\cG}[2]{\makecell[c]{#1 \\[-2pt] \textcolor{teal!90!black}{\fontsize{7pt}{8pt}\selectfont(+#2)}}}
\newcommand{\cBD}[2]{\makecell[c]{#1 \\[-2pt] \textbf{\textcolor{red}{\fontsize{7pt}{8pt}\selectfont(#2)}}}}
\newcommand{\cBG}[2]{\makecell[c]{#1 \\[-2pt] \textbf{\textcolor{teal}{\fontsize{7pt}{8pt}\selectfont(+#2)}}}}

\setlength{\tabcolsep}{2.5pt} 
\renewcommand{\arraystretch}{1.45}

\resizebox{\textwidth}{!}{%
\begin{tabular}{@{} l l cccc @{\hspace{5pt}} cccc @{\hspace{5pt}} cccc @{\hspace{5pt}} cccc @{}}
\toprule
\multirow{2.5}{*}{\textbf{Model}} & \multirow{2.5}{*}{\textbf{Instruction}} 
& \multicolumn{4}{c@{\hspace{5pt}}}{\makecell{\textbf{AV-Critical} \\ \textbf{(35.4\%)}}} 
& \multicolumn{4}{c@{\hspace{5pt}}}{\makecell{\textbf{AV-Supportive} \\ \textbf{(38.4\%)}}} 
& \multicolumn{4}{c@{\hspace{5pt}}}{\makecell{\textbf{AV-Present} \\ \textbf{(26.2\%)}}} 
& \multicolumn{4}{c}{\makecell{\textbf{Overall} \\ \textbf{(100\%)}}} \\
\cmidrule(lr){3-6} \cmidrule(lr){7-10} \cmidrule(lr){11-14} \cmidrule(l){15-18}
& & \textbf{TM} & \textbf{EM} & \textbf{SR} & \textbf{GP} 
  & \textbf{TM} & \textbf{EM} & \textbf{SR} & \textbf{GP} 
  & \textbf{TM} & \textbf{EM} & \textbf{SR} & \textbf{GP} 
  & \textbf{TM} & \textbf{EM} & \textbf{SR} & \textbf{GP} \\
\midrule

\multirow{2}{*}{Gemini 3 Pro} 
& Text (Baseline) & 76.9 & 57.9 & 33.2 & 42.2 & 79.6 & 65.2 & 34.4 & 45.0 & 84.7 & 69.0 & 33.0 & 44.4 & 80.0 & 63.6 & 33.4 & 43.6 \\
& TTS Voice       & \cBD{69.0}{-7.9} & \cBD{52.1}{-5.8} & \cD{29.1}{-4.1} & \cD{39.4}{-2.8} & \cD{74.5}{-5.1} & \cD{59.9}{-5.3} & \cBD{27.7}{-6.7} & \cBD{40.3}{-4.7} & \cD{81.6}{-3.1} & \cD{67.3}{-1.7} & \cBG{35.8}{2.8} & \cBG{46.4}{2.0} & \cD{74.3}{-5.7} & \cD{59.1}{-4.5} & \cD{30.3}{-3.1} & \cD{41.6}{-2.0} \\

\midrule

\multirow{2}{*}{Qwen3-Omni} 
& Text (Baseline) & 58.0 & 29.4 & 7.0 & 18.5 & 65.1 & 33.5 & 4.8 & 17.4 & 66.9 & 34.4 & 3.2 & 16.0 & 63.1 & 32.3 & 5.1 & 17.4 \\
& TTS Voice       & \cD{55.7}{-2.3} & \cD{26.6}{-2.8} & \cD{6.0}{-1.0} & \cD{16.8}{-1.7} & \cBD{58.8}{-6.3} & \cBD{27.2}{-6.3} & \cBD{1.8}{-3.0} & \cBD{14.7}{-2.7} & \cD{62.6}{-4.3} & \cD{33.9}{-1.5} & \cBG{3.7}{0.5} & \cBG{17.7}{1.7} & \cD{58.7}{-4.4} & \cD{28.5}{-3.8} & \cD{3.8}{-1.3} & \cD{16.2}{-1.2} \\

\bottomrule
\end{tabular}%
}
\vspace{-3mm}
\end{table*}

\begin{figure*}[t]
\centering
\includegraphics[width=\textwidth]{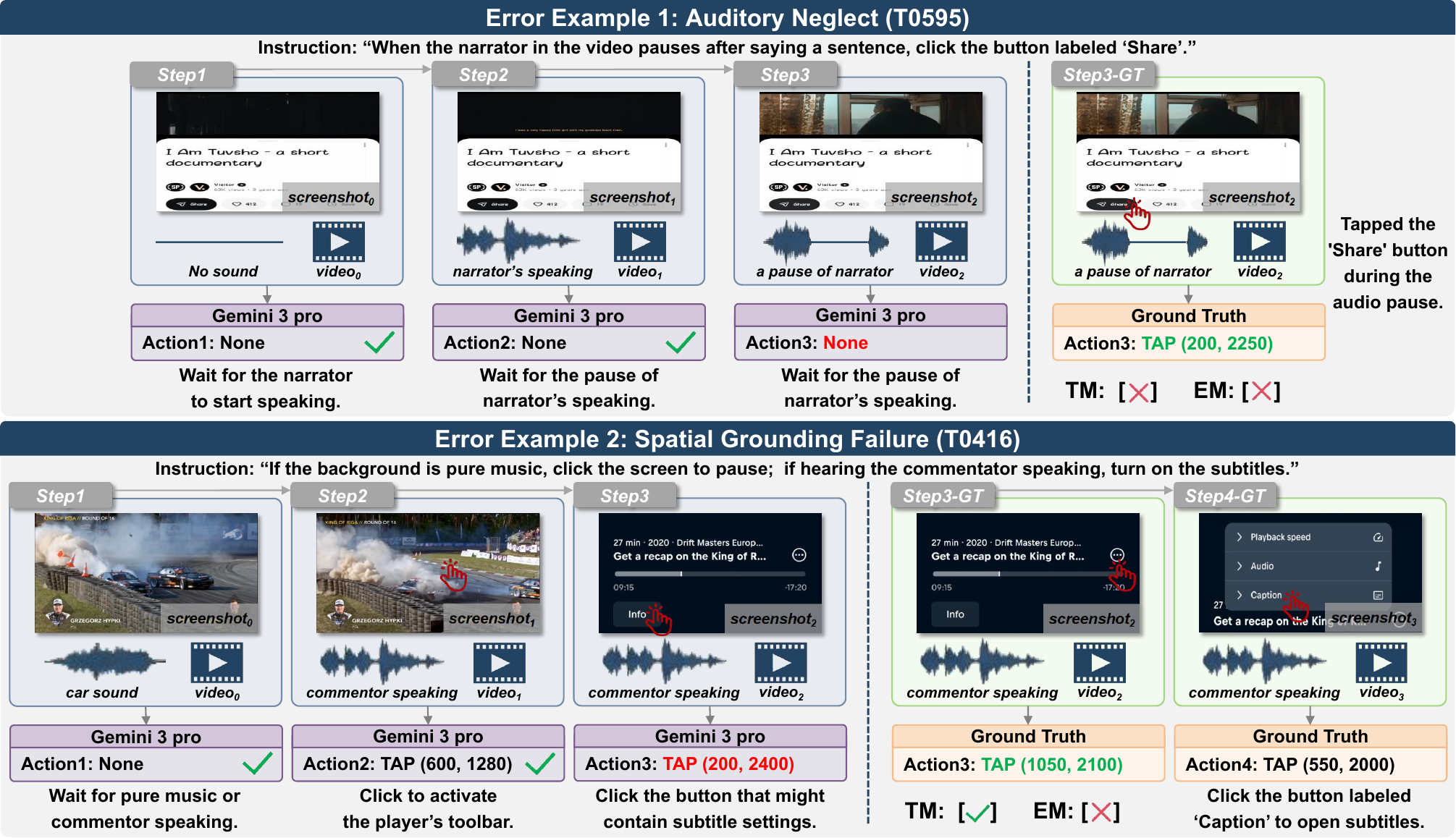}
\caption{
\textbf{Qualitative error analysis of Gemini 3.0 Pro.} 
(Top) \textit{Auditory Neglect}: The model fails to trigger an action in response to a transient acoustic state change (a pause in narration). 
(Bottom) \textit{Spatial Grounding Failure}: The model correctly identifies the target action primitive (\texttt{TAP}) based on the multimodal context but fails to predict the precise spatial coordinates of the subtitle icon.
}
\label{fig:error_cases}
\end{figure*}

\subsection{Error Analysis}
\label{sec:error_analysis}

To further investigate the operational bottlenecks of current models in omni-modal environments, we perform a qualitative analysis on representative failure cases from Gemini 3.0 Pro. Based on empirical observations of the predicted trajectories, we highlight two recurring error patterns that explicitly illustrate the difficulties of multimodal grounding in GUI tasks.

\paragraph{Auditory Neglect.} 
Figure~\ref{fig:error_cases} (top) illustrates a failure on task T4014 (within the Vimeo app), where the execution timing depends strictly on a transient audio event. The instruction requires the agent to tap the ``Share'' button specifically when the video narrator pauses. At Step 1 and Step 2, the model correctly outputs \texttt{NONE} while the audio contains silence and continuous speech, respectively. However, at Step 3, when the required audio pause occurs, the model continues to predict \texttt{NONE} instead of the ground-truth \texttt{TAP} action. This case demonstrates an instance where the model fails to map a step-level acoustic state change to the corresponding action execution, resulting in both Type Match (TM) and Exact Match (EM) failures.

\paragraph{Spatial Grounding Failure.} 
Figure~\ref{fig:error_cases} (bottom) depicts task T4210 (within the Red Bull TV app), which requires invoking the video toolbar and enabling subtitles upon hearing the commentator. At Step 2, the model correctly predicts a \texttt{TAP} to activate the toolbar. At Step 3, the model correctly predicts the action type (\texttt{TAP}) to interact with the subtitle settings, successfully fulfilling the TM metric. However, the predicted coordinates $(200, 2400)$ deviate significantly from the ground-truth bounding box of the subtitle icon $(1050, 2100)$. This isolated EM failure indicates that while the model successfully comprehends the multimodal instruction and determines the correct operational primitive, precise spatial grounding on the complex visual interface remains a challenge.

\section{Conclusion and Future Work}
\label{sec:conclusion}

We introduced OmniGUI, the first step-level benchmark designed to evaluate GUI agents in omni-modal smartphone environments. Unlike prior benchmarks relying on static screenshots, OmniGUI provides continuous, interleaved multimodal inputs---comprising static images, temporal video clips, and synchronous audio---at every action step. The benchmark encompasses 709 expert-demonstrated episodes (2,579 steps) systematically distributed across five cognitive dimensions and three objective multimodal dependency levels.

Our extensive evaluation establishes initial baselines by utilizing foundational omni-modal models as agent proxies. The empirical results demonstrate that while current models exhibit competency on static visual tasks, their action prediction performance degrades significantly in environments requiring synchronous temporal and auditory signals. Furthermore, our ablation studies isolate specific operational bottlenecks, such as cross-modal interference and performance degradation when processing dual-audio streams. 

While OmniGUI establishes a comprehensive evaluation foundation, our current protocol employs an offline, step-level methodology using expert-demonstrated action histories. While this optimally isolates per-step perception-to-action capabilities and ensures deterministic reproducibility, it does not evaluate an agent's ability to recover from compounding errors during autonomous, end-to-end rollouts. Future work could explore extending the evaluation protocol to include autonomous, interactive settings to assess these dynamic error-recovery mechanisms.

%
%
\bibliographystyle{splncs04}
\bibliography{main}

\appendix

\section{Experimental Details}
\label{sec:appendix_experimental_details}

This section details the exact prompt templates, user message structures, and hyperparameter configurations used in our evaluations. All evaluated models share an identical prompt structure without any model-specific prompt engineering.

\subsection{Unified Prompt Templates and Input Structure}
\label{sec:appendix_prompt}

\newtcolorbox{promptbox}[1]{
  enhanced,
  breakable,
  colback=gray!5,           
  colframe=teal!80!black,   
  colbacktitle=teal!80!black,
  title={\textbf{\textsf{#1}}}, 
  fonttitle=\normalsize,
  arc=3pt,                  
  boxrule=1pt,
  toprule at break=1pt,      
  bottomrule at break=1pt,   
  pad at break=2mm,          
  left=8pt, right=8pt, top=8pt, bottom=8pt,
  boxsep=0pt,
  toptitle=4pt, bottomtitle=4pt
}

The baseline evaluation utilizes a standardized two-part prompt structure: a System Prompt and an interleaved User Message.

\vspace{3mm}
\begin{promptbox}{System Prompt}
\scriptsize \ttfamily
You are an Android GUI Agent. You are given a task and your action history, \\
with screenshots, audio and video. You need to perform the next action to \\
complete the task. \\

\#\# Output Format \\
Output ONLY a single JSON object. No markdown, no code blocks, no explanation. \\
Your ENTIRE response must be valid JSON and nothing else. \\
You MUST ALWAYS output exactly one JSON object. NEVER output an empty response. \\

\#\# Action Space \\
| action\_type | Name \hspace{12.5mm} | Parameters | \\
|-------------|------------------|--------------------------------------------------| \\
| -1 \hspace{8mm} | NONE \hspace{12.5mm} | (none) - Wait/observe without action \hspace{13mm} | \\
| 0 \hspace{9.5mm} | TAP \hspace{13.5mm} | coordinate: [x, y] \hspace{31mm} | \\
| 1 \hspace{9.5mm} | DOUBLE\_TAP \hspace{6mm} | coordinate: [x, y] \hspace{31mm} | \\
| 2 \hspace{9.5mm} | LONG\_PRESS \hspace{6mm} | coordinate:[x, y] \hspace{31mm} | \\
| 3 \hspace{9.5mm} | SWIPE\_UP \hspace{8mm} | start\_coordinate: [x, y], end\_coordinate: [x, y] | \\
| 4 \hspace{9.5mm} | SWIPE\_DOWN \hspace{6mm} | start\_coordinate: [x, y], end\_coordinate:[x, y] | \\
| 5 \hspace{9.5mm} | SWIPE\_LEFT \hspace{6mm} | start\_coordinate: [x, y], end\_coordinate: [x, y] | \\
| 6 \hspace{9.5mm} | SWIPE\_RIGHT \hspace{4.5mm}| start\_coordinate: [x, y], end\_coordinate: [x, y] | \\
| 7 \hspace{9.5mm} | INPUT \hspace{11.5mm} | text: "string" \hspace{35mm} | \\
| 8 \hspace{9.5mm} | BACK \hspace{12.5mm} | (none) \hspace{43.5mm} | \\
| 9 \hspace{9.5mm} | HOME \hspace{12.5mm} | (none) \hspace{43.5mm} | \\
| 10 \hspace{8mm} | TASK\_COMPLETE \hspace{2.5mm}| (none) \hspace{43.5mm} | \\
| 11 \hspace{8mm} | TASK\_IMPOSSIBLE | (none) \hspace{43.5mm} | \\

\#\# Coordinate System \\
- Format:[x, y] where x=horizontal (left->right), y=vertical (top->bottom) \\
- Range: 0-1000 (normalized).[0, 0] = top-left,[1000, 1000] = bottom-right. \\

\#\# Rules \\
- Base actions on the current screenshot (the last image in your input). \\
- If the task appears COMPLETE or already done, you MUST output \{"action\_type": 10\}. \\
- If the task requires waiting/observing, output \{"action\_type": -1\}. \\
- If the task is impossible, output \{"action\_type": 11\}. \\
- Even when unsure, you MUST still output a valid JSON action. An empty response \\
\hspace*{2mm} is NEVER acceptable. \\

\#\# Examples \\
- Wait/observe: \{"action\_type": -1\} \\
- Tap: \{"action\_type": 0, "coordinate": [500, 500]\} \\
- Swipe down: \{"action\_type": 4, "start\_coordinate":[500, 300], \\
\hspace*{2mm} "end\_coordinate": [500, 700]\} \\
- Input text: \{"action\_type": 7, "text": "hello"\} \\
- Back: \{"action\_type": 8\} \\
- Task done: \{"action\_type": 10\}
\end{promptbox}

\vspace{3mm}
The \textbf{User Message} strictly follows an interleaved multimodal sequence, combining historical visual context with current-step multimodal signals. To maintain ecological validity, the \texttt{\{task\_description\}} dynamically injects either the English or Chinese instruction corresponding to the target application's native environment.

\vspace{3mm}
\begin{promptbox}{User Message Structure}
\scriptsize \ttfamily
{[1]} (optional) Text:  "[History Screenshot]:" \\
{[2]} (optional) Image: <history screenshot from step t-2> \\
{[3]} (optional) Video: <current step video clip> \\
{[4]} (optional) Audio: <current step environment audio> \\
{[5]} \hspace{11.5mm} Image: <current screenshot> \\
{[6]} \hspace{11.5mm} Text:  <prompt text --- see below> \\

--- Prompt Text --- \\
Goal: \{task\_description\} \\

Step: \{current\_step\_index\} \\

Action History: \\
Step 0: \{action\_text\_0\} \\
Step 1: \{action\_text\_1\} \\
... \\
Step N: \{action\_text\_N\} \\

Analyze the current screenshot and output the next action as JSON.
\end{promptbox}

\vspace{3mm}

\paragraph{Ablation Configurations.} 
For the ablation experiments, the input structures are deterministically modified to control specific variables. These modifications are strictly applied at the input level.

For the modality ablation experiments (Section 4.3), the respective media payloads (audio, video, or both) are physically omitted from the User Message. Concurrently, the first sentence of the System Prompt is minimally adjusted to reflect the available modalities:

\vspace{2mm}
\begin{promptbox}{System Prompt Modifications (Modality Ablation)}
\scriptsize \ttfamily[w/o Audio]
You are an Android GUI Agent. You are given a task and your action history, \\
with screenshots and video. You need to perform the next action...

[w/o Video]
You are an Android GUI Agent. You are given a task and your action history, \\
with screenshots and audio. You need to perform the next action...

[w/o Audio \& Video]
You are an Android GUI Agent. You are given a task and your action history, \\
with screenshots. You need to perform the next action...
\end{promptbox}
\vspace{2mm}

To evaluate cognitive load during dual-audio processing (Text vs. TTS Voice), a \texttt{.wav} audio file containing the spoken instruction is injected into the User Message immediately preceding the text prompt. The textual \texttt{\{task\_description\}} is replaced by a static placeholder, and the System Prompt is modified to include a listening directive:

\vspace{2mm}
\begin{promptbox}{Prompt Modifications (TTS Voice Instruction)}
\scriptsize \ttfamily
[System Prompt Adjustment]
Listen to the voice instruction audio for the task description. \\
You are an Android GUI Agent. You are given a task and your action history, \\
with screenshots, audio and video. You need to perform the next action...

[User Message Adjustment]
... \\
{[4]} (optional) Audio: <current step environment audio> \\
{[4.1]}          Audio: <TTS spoken instruction .wav file> \\
{[5]} \hspace{11.5mm} Image: <current screenshot> \\
{[6]} \hspace{11.5mm} Text:  <prompt text --- see below> \\

--- Prompt Text --- \\
Goal: [Please listen to the voice instruction] \\

Step: \{current\_step\_index\} \\
...
\end{promptbox}

\subsection{Model Configurations and Hyperparameters}
\label{sec:appendix_hyperparams}

To ensure deterministic, reproducible outputs and to evaluate the models' most confident decision boundaries, we enforced greedy decoding strategies across all evaluations. Table~\ref{tab:hyperparameters} details the generation hyperparameters used for each model. For proprietary API models, parameters were explicitly set to zero where supported. For certain open-source models deployed via default server interfaces (e.g., VITA, Baichuan-Omni-1.5), evaluations were conducted strictly under their officially recommended deterministic inference configurations.

\begin{table}[h]
\centering
\caption{Hyperparameter configurations for all evaluated models.}
\label{tab:hyperparameters}
\renewcommand{\arraystretch}{1.15}
\resizebox{\textwidth}{!}{%
\begin{tabular}{l c c c c c c}
\toprule
\textbf{Model} & \textbf{Temperature} & \textbf{Max Tokens} & \textbf{Top\_p} & \textbf{Top\_k} & \textbf{Do\_Sample} & \textbf{Seed} \\
\midrule
Gemini 3.0 Pro & 0.0 & 4096 & - & - & - & - \\
Gemini 3.0 Flash & 0.0 & 4096 & - & - & - & - \\
Gemini 2.5 Pro & 0.0 & 4096 & - & - & - & - \\
Gemini 2.5 Flash & 0.0 & 4096 & - & - & - & - \\
Qwen3-Omni & 0.0 & 4096 & - & - & - & - \\
MiniCPM-o 4.5 & 0.0 & 4096 & - & - & \texttt{False} & - \\
VITA-1.5 & Server Default & Server Default & - & - & - & - \\
Baichuan-Omni-1.5 & Server Default & Server Default & - & - & - & - \\
\bottomrule
\end{tabular}%
}
\end{table}

\section{Dataset Construction Details}
\label{sec:appendix_dataset}

This section provides detailed dataset statistics for the OmniGUI benchmark. Table~\ref{tab:full_app_breakdown} presents a comprehensive application-level breakdown, detailing the exact distribution of episodes and action steps across the five task dimensions and the three multimodal dependency levels for each of the 29 evaluated smartphone applications.

\begin{table*}[htbp]
\centering
\caption{
\textbf{Comprehensive breakdown of the OmniGUI dataset per application.} 
The table reports the volume of episodes (Ep.) and total steps (Stp.) alongside their exact distribution across the five Task Dimensions (\textit{Loc.} = Localization, \textit{Sem.} = Semantic Understanding, \textit{CrM.} = Cross-modal Discrimination, \textit{Tmp.} = Temporal Reasoning, \textit{Ins.} = Instant Response) and the three Multimodal Dependency levels (\textit{Cri.} = AV-Critical, \textit{Sup.} = AV-Supportive, \textit{Pre.} = AV-Present).
}
\label{tab:full_app_breakdown}
\vspace{-2mm}
\setlength{\tabcolsep}{3.5pt} 
\renewcommand{\arraystretch}{1.15} 
\resizebox{\textwidth}{!}{%
\begin{tabular}{@{} l cc ccccc ccc @{\hspace{15pt}} l cc ccccc ccc @{}}
\toprule
\multicolumn{11}{c@{\hspace{15pt}}}{\textbf{Chinese Applications (ZH)}} & \multicolumn{11}{c}{\textbf{English Applications (EN)}} \\
\cmidrule(r){1-11} \cmidrule(l){12-22}
\multirow{2}{*}{\textbf{App Name}} & \multirow{2}{*}{\textbf{Ep.}} & \multirow{2}{*}{\textbf{Stp.}} & \multicolumn{5}{c}{\textbf{Task Dimensions}} & \multicolumn{3}{c}{\textbf{Modality Dep.}} & \multirow{2}{*}{\textbf{App Name}} & \multirow{2}{*}{\textbf{Ep.}} & \multirow{2}{*}{\textbf{Stp.}} & \multicolumn{5}{c}{\textbf{Task Dimensions}} & \multicolumn{3}{c}{\textbf{Modality Dep.}} \\
\cmidrule(lr){4-8} \cmidrule(lr){9-11} \cmidrule(lr){15-19} \cmidrule(l){20-22}
& & & \textbf{Loc} & \textbf{Sem} & \textbf{CrM} & \textbf{Tmp} & \textbf{Ins} & \textbf{Cri} & \textbf{Sup} & \textbf{Pre} & & & & \textbf{Loc} & \textbf{Sem} & \textbf{CrM} & \textbf{Tmp} & \textbf{Ins} & \textbf{Cri} & \textbf{Sup} & \textbf{Pre} \\
\midrule

Bilibili    & 29  & 91  & 7 & 5 & 6 & 6 & 5 & 10 & 15 & 4  & Duolingo    & 35  & 133 & 5 & 5 & 5 & 15 & 5 & 16 & 11 & 8 \\
Douyin      & 29  & 97  & 7 & 5 & 6 & 6 & 5 & 12 & 8  & 9  & Vimeo       & 25  & 86  & 5 & 5 & 5 & 5  & 5 & 12 & 6  & 7 \\
Meituan     & 26  & 103 & 6 & 5 & 6 & 5 & 4 & 4  & 5  & 17 & TED         & 25  & 115 & 5 & 5 & 5 & 5  & 5 & 16 & 6  & 3 \\
QQ Music    & 25  & 95  & 5 & 5 & 5 & 5 & 5 & 1  & 11 & 13 & Snapchat    & 25  & 104 & 5 & 5 & 5 & 5  & 5 & 8  & 11 & 6 \\
PDD         & 25  & 72  & 5 & 5 & 5 & 5 & 5 & 5  & 9  & 11 & Spotify     & 25  & 91  & 5 & 5 & 5 & 5  & 5 & 7  & 12 & 6 \\
DiDi        & 25  & 92  & 5 & 5 & 5 & 5 & 5 & 2  & 5  & 18 & Tasty       & 25  & 69  & 5 & 5 & 10& 5  & 0 & 18 & 5  & 2 \\
JD          & 25  & 80  & 5 & 5 & 5 & 5 & 5 & 3  & 9  & 13 & GTrans      & 25  & 88  & 5 & 5 & 5 & 5  & 5 & 11 & 14 & 0 \\
Weibo       & 25  & 106 & 5 & 5 & 5 & 5 & 5 & 7  & 12 & 6  & X           & 25  & 61  & 5 & 5 & 5 & 5  & 5 & 5  & 8  & 12\\
WeChat      & 25  & 84  & 7 & 5 & 4 & 5 & 4 & 9  & 8  & 8  & TikTok      & 25  & 77  & 5 & 5 & 5 & 5  & 5 & 1  & 9  & 15\\
Amap        & 25  & 64  & 5 & 5 & 5 & 5 & 5 & 1  & 3  & 21 & YouTube     & 25  & 123 & 5 & 5 & 5 & 5  & 5 & 3  & 12 & 10\\
RedNote     & 25  & 99  & 5 & 5 & 5 & 5 & 5 & 11 & 6  & 8  & RedBull     & 24  & 110 & 5 & 4 & 5 & 5  & 5 & 14 & 5  & 5 \\
Kuaishou    & 24  & 122 & 4 & 5 & 5 & 6 & 4 & 11 & 6  & 7  & Amazon      & 22  & 69  & 5 & 5 & 4 & 4  & 4 & 1  & 4  & 17\\
Taobao      & 21  & 78  & 5 & 5 & 2 & 4 & 5 & 2  & 4  & 15 & IMDb        & 20  & 95  & 5 & 4 & 4 & 3  & 4 & 10 & 9  & 1 \\
iQIYI       & 20  & 56  & 0 & 5 & 0 & 12& 3 & 2  & 7  & 11 & Insta       & 20  & 55  & 4 & 4 & 4 & 1  & 7 & 5  & 9  & 6 \\
Alipay      & 14  & 64  & 5 & 0 & 5 & 4 & 0 & 4  & 1  & 9  & \multicolumn{11}{c}{} \\

\midrule
\multicolumn{22}{l}{\textbf{TOTAL (ALL 29 APPS):} \textbf{709 Episodes, 2579 Steps} $\mid$ Task Dims: Loc(145), Sem(137), CrM(141), Tmp(156), Ins(130) $\mid$ Dep: Cri(211), Sup(230), Pre(268)} \\
\bottomrule
\end{tabular}%
}
\end{table*}

\subsection{Data Format Example}
\label{sec:appendix_data_format}

The OmniGUI dataset is hierarchically organized by application to facilitate structured access and reproducible evaluations. For each application, the dataset separates episode-level metadata from step-level multimodal assets and execution traces. 

\paragraph{1. Directory Structure.} 
Listing~\ref{lst:dir_tree} illustrates the standard directory hierarchy for a given application (e.g., \texttt{TED}). The episode-level metadata is distributed across five \texttt{.jsonl} files, corresponding to the cognitive task dimensions. The \texttt{media} directory encapsulates individual episode folders, which store the atomic step-level data including interleaved video clips, audio tracks, screenshots, and the step-wise action trace JSON files. The \texttt{dataset\_TED.json} file serves as a global index, compiling all step-level annotations into a single array to streamline batch dataloading; its internal schema is identical to the step-level traces detailed in Listing~\ref{lst:json_example}.

\newtcolorbox{codebox}[1][]{
  enhanced, breakable,
  colback=blue!4,           
  colframe=blue!60!black,   
  title={#1}, fonttitle=\bfseries\sffamily\small,
  boxrule=0.8pt, arc=2pt,
  left=6pt, right=6pt, top=4pt, bottom=4pt,
  toptitle=3pt, bottomtitle=3pt,
  toprule at break=0.8pt, bottomrule at break=0.8pt
}

\vspace{2mm}
\begin{codebox}[Listing 1: Standardized application directory hierarchy \label{lst:dir_tree}]
\scriptsize
\begin{verbatim}
[App_Name]/ (e.g., TED/)
|-- localization.jsonl
|-- semantic_understanding.jsonl
|-- cross_modal_discrimination.jsonl
|-- temporal_reasoning.jsonl
|-- instant_response.jsonl
|-- dataset_TED.json  // Global index aggregating all steps
\-- media/
    |-- T009/
    |   |-- 1.mp4           // Step 1 video clip
    |   |-- 1.wav           // Step 1 audio track
    |   |-- 1.png           // Step 1 observation screenshot
    |   |-- 2.mp4
    |   |-- ...
    |   \-- T009.json       // Step-level action trace array
    |-- T019/
    \-- ...
\end{verbatim}
\end{codebox}
\vspace{2mm}

\paragraph{2. Episode-Level Metadata (.jsonl).} 
The \texttt{.jsonl} files store the task descriptions for each episode. Listing~\ref{lst:jsonl_example} presents a snippet from the \texttt{TED} application. These bilingual instructions map directly to the \texttt{\{task\_description\}} placeholder in the unified evaluation prompt.

\vspace{2mm}
\begin{codebox}[Listing 2: Episode metadata format (Snippet from temporal\_reasoning.jsonl) \label{lst:jsonl_example}]
\scriptsize
\begin{CJK*}{UTF8}{gbsn}
\raggedright
\noindent\texttt{\{"ID": "T0547", "app": "TED", "instruction\_zh": "}%
视频开始播放标志性的红白‘水滴/宇宙’展开动画，\\
\noindent\hspace*{1.5em}当动画瞬间结束切入真实舞台场景的时刻，立即点击暂停视频。\texttt{",}\\
\noindent\texttt{ "instruction\_en": "As the video starts with the iconic red and white}\\
\noindent\texttt{ drop/universe animation, tap pause the moment the animation ends."\}}\\[2pt]

\noindent\texttt{\{"ID": "T0548", "app": "TED", "instruction\_zh": "}%
观察画面与声音，当第二位主讲人开口发声的\\
\noindent\hspace*{1.5em}一瞬间，立刻点击暂停视频。\texttt{",}\\
\noindent\texttt{ "instruction\_en": "Observe the video and audio, and tap pause the instant}\\
\noindent\texttt{ the second speaker begins to speak."\}}\\[2pt]

\noindent\texttt{\{"ID": "T0551", "app": "TED", "instruction\_zh": "}%
点击首页下方正在播放的视频，当视频瞬间放大进入全屏播放的时刻，\\
\noindent\hspace*{1.5em}立即下滑退出视频。\texttt{",}\\
\noindent\texttt{ "instruction\_en": "Tap the currently playing video at the bottom of the Home screen, and}\\
\noindent\texttt{ the moment it expands to full screen, swipe down immediately to exit."\}}
\end{CJK*}
\end{codebox}
\vspace{2mm}

\paragraph{3. Step-Level Execution Trace (.json).} 
Inside each episode's media folder, a specific JSON file logs the chronological action sequence. Listing~\ref{lst:json_example} presents a snippet from an AV-Critical episode, demonstrating how waiting states (\texttt{NONE}) and execution states (\texttt{TAP}) are recorded alongside precise target bounding boxes.

\vspace{2mm}

\begin{codebox}[Listing 3: Step-level execution trace (Snippet from T3609.json) \label{lst:json_example}]
\scriptsize
\begin{CJK*}{UTF8}{gbsn}
\raggedright
\noindent\texttt{// 1. Episode Metadata (from JSONL)}\\
\noindent\texttt{\{}\\
\noindent\texttt{  "ID": "T0540",}\\
\noindent\texttt{  "app": "TED",}\\
\noindent\texttt{  "instruction\_zh": "}%
闭屏听声音。如果听到‘观众大笑’则点击‘Share（分享）’按钮，\\
\noindent\hspace*{1.5em}如听到‘鼓掌’，则点击‘like（喜欢）’按钮,\texttt{",}\\
\noindent\texttt{  "instruction\_en": "Listen to the audio with the screen off. If you hear}\\
\noindent\texttt{  "Laughter", tap the "Share" button; if you hear "Applause", tap "Like"."}\\
\noindent\texttt{\}}\\[2pt]
\end{CJK*}

\begin{verbatim}
// 2. Step-level Array (Truncated for display)[
  {
    "episode_id": "T0540",
    "episode_length": 7,
    "step_id": 0,
    "instruction_en": "Listen to the audio with the screen off...",
    "video_path": "T0540/1.mp4",
    "audio_path": "T0540/1.wav",
    "image_path": "T0540/1.png",
    "image_width": 720,
    "image_height": 1600,
    "result_action_type": -1,       // Mapped to Action Space: NONE (Wait)
    "result_action_text": "",
    "result_touch_xy": "",
    "result_lift_xy": "",
    "result_action_json": "{\"action_type\":-1,\"text\":\"\",\"touch_xy\":\"\"}"
  },
  ... // Steps 1 to 4 omitted (waiting states with continuous audio input)
  {
    "episode_id": "T0540",
    "episode_length": 7,
    "step_id": 5,
    "instruction_en": "Listen to the audio with the screen off...",
    "video_path": "T0540/6.mp4",
    "audio_path": "T0540/6.wav",
    "image_path": "T0540/6.png",
    "image_width": 720,
    "image_height": 1600,
    "result_action_type": 0,        // Mapped to Action Space: TAP
    "result_action_text": "",
    "result_touch_xy": "[[210,786],[351,850]]", // Target bounding box
    "result_lift_xy": "",
    // Long string manually wrapped for display formatting
    "result_action_json": "{\"action_type\":0,\"text\":\"\",
                           \"touch_xy\":\"[[210,786],[351,850]]\"}"
  }
]
\end{verbatim}
\end{codebox}

\section{Additional Experimental Results}
\label{sec:appendix_additional_results}

This section provides supplementary visual analyses of the baseline evaluations, offering a more granular perspective on model capabilities across different cognitive dimensions, multimodal dependencies, and application environments.

\subsection{Model Capability Fingerprints}
\label{sec:appendix_radar}

\begin{figure}[htbp]
\centering
\begin{minipage}{0.47\textwidth}
  \centering
  \includegraphics[width=\linewidth]{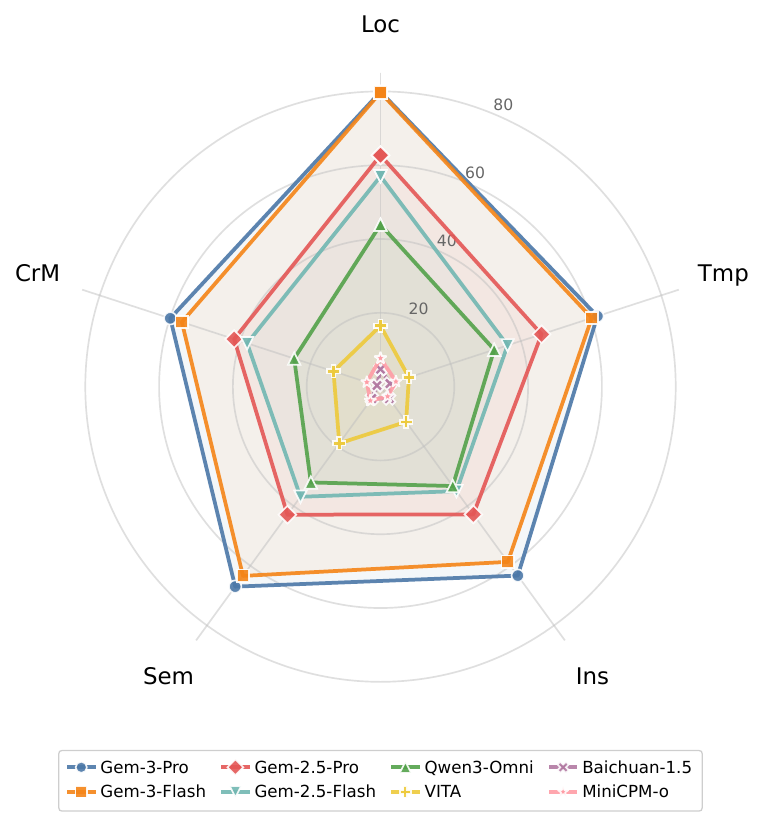}
  \centerline{(a) Performance across 5 Task Dimensions}
\end{minipage}\hfill
\begin{minipage}{0.495\textwidth}
  \centering
  \includegraphics[width=\linewidth]{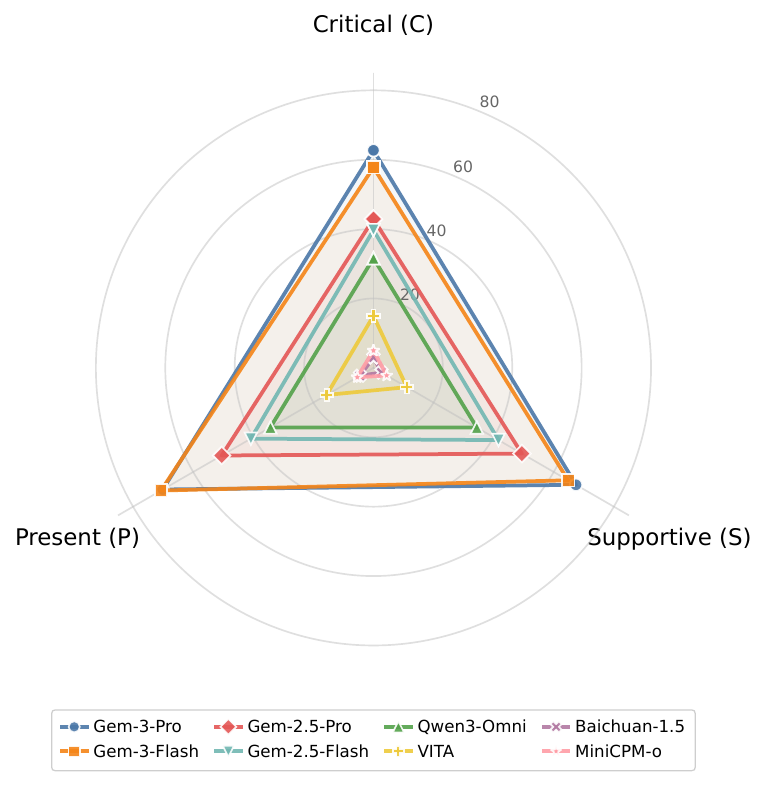}
  \centerline{(b) Performance across 3 Dependency Levels}
\end{minipage}
\caption{
\textbf{Capability fingerprints of evaluated models.} The radar charts map the Exact Match (EM) performance. (a) Across the five operational dimensions, models universally exhibit stronger capabilities on static Localization (Loc) compared to Temporal Reasoning (Tmp) and Cross-modal Discrimination (CrM). (b) Across the multimodal dependency levels, performance contracts monotonically as tasks transition from AV-Present (P) to AV-Critical (C), visually validating the necessity of multimodal perception mechanisms.
}
\label{fig:radar_charts}
\end{figure}

To visualize the specific performance profiles of the evaluated models, Figure~\ref{fig:radar_charts} presents multidimensional capability fingerprints using the Exact Match (EM) metric. The analysis is divided into two perspectives: performance across the five predefined task dimensions and performance across the three multimodal dependency levels.

\subsection{Performance Breakdown by Application}
\label{sec:appendix_app_breakdown}

\begin{figure*}[htbp]
\centering
\includegraphics[width=\textwidth]{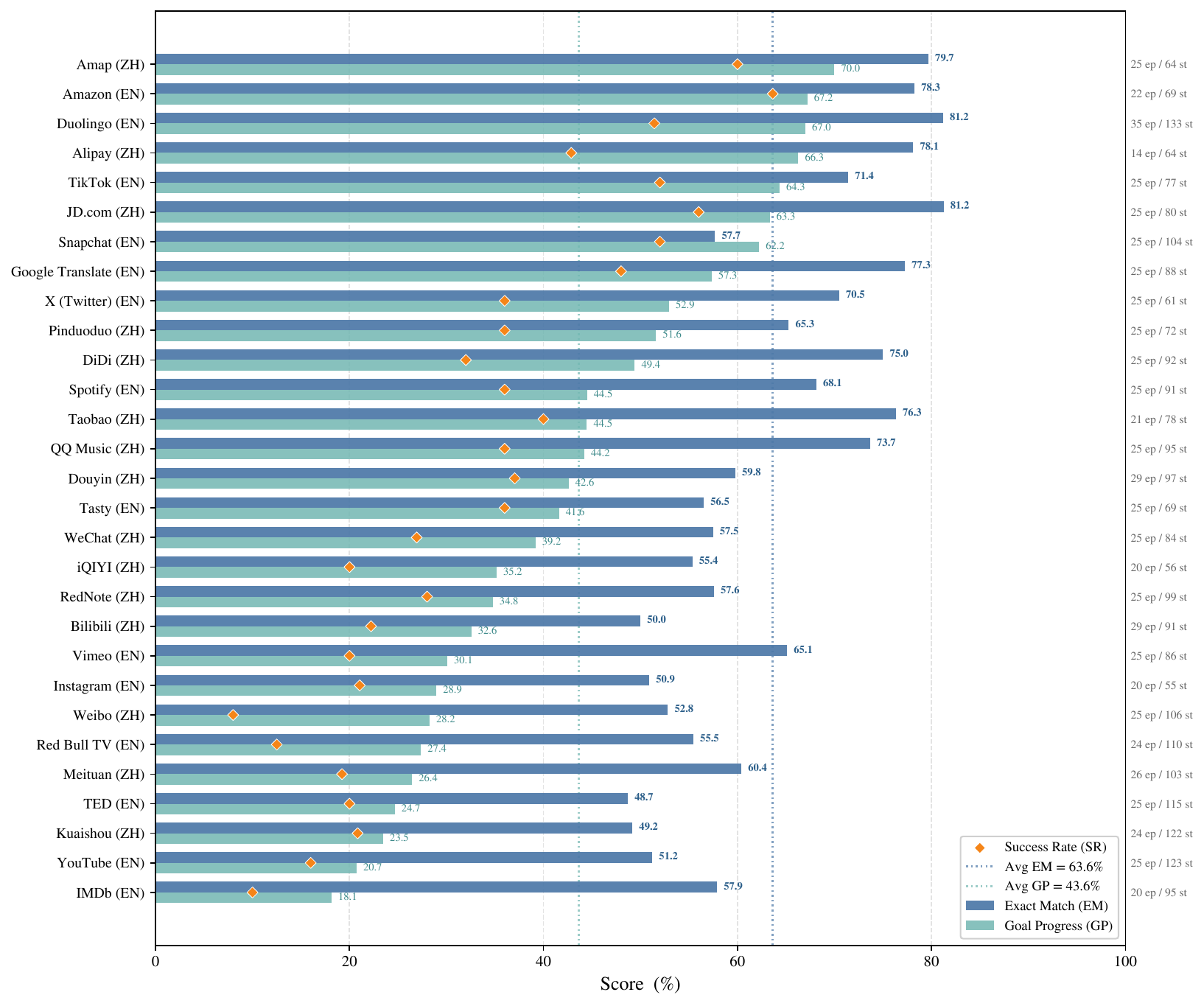}
\caption{
\textbf{Gemini 3.0 Pro performance disaggregated by application.} The applications are sorted by Exact Match (EM) scores. The right-hand axis explicitly denotes the sample volume (episodes/steps) for each application. The vertical dotted lines represent the overall benchmark averages, providing a reference to identify application environments that present above-average or below-average difficulty.
}
\label{fig:app_breakdown}
\end{figure*}

To illustrate the variance in execution difficulty across different smartphone interfaces, Figure~\ref{fig:app_breakdown} details the performance of the strongest baseline model, Gemini 3.0 Pro, disaggregated by the 29 evaluated applications. 

The horizontal bar chart reports the Exact Match (EM), Goal Progress (GP), and Success Rate (SR) metrics, sorted in ascending order by EM from bottom to top. The overall benchmark averages (\text{EM} = 63.6\%, \text{GP} = 43.6\%) are indicated by vertical dotted lines. The variance across applications objectively underscores the diversity of GUI complexities captured within OmniGUI.

\end{document}